\documentclass[
  journal=pasa,
  manuscript=research-paper,
  year=2025,
  volume=YY,
]{cup-journal}

\usepackage{aas-macros}
\received {dd Mmm YYYY}
\revised  {dd Mmm YYYY}
\accepted {05 Nov 2025}
\published{dd Mmm YYYY}

\usepackage{amsmath}
\usepackage[nopatch]{microtype}
\usepackage{booktabs}
\usepackage{graphicx}
\usepackage{amsmath}
\usepackage{epstopdf}
\usepackage{subcaption}
\usepackage{caption}
\usepackage{rotating}
\usepackage{pdflscape}
\usepackage{flafter}
\usepackage{afterpage}
\usepackage{lipsum}
\usepackage{float}
\restylefloat{table}
\usepackage{placeins}
\usepackage{longtable}
\newcommand{\myedit}{}
\usepackage{tabularx}
\newcommand\setrow[1]{\gdef\rowmac{#1}#1\ignorespaces}
\newcommand\clearrow{\global\let\rowmac\relax}
\clearrow

\title{An investigation into correlations between FRB and host galaxy properties}
\author{M. Glowacki}
\affiliation{Institute for Astronomy, University of Edinburgh, Royal Observatory, Edinburgh, EH9 3HJ, United Kingdom}
\alsoaffiliation{International Centre for Radio Astronomy Research, Curtin University, Bentley, 6102, WA, Australia}
\alsoaffiliation{Inter-University Institute for Data Intensive Astronomy, Department of Astronomy, University of Cape Town, Cape Town, South Africa}
\email[M. Glowacki]{marcin.glowacki@roe.ac.uk}

\author{A. Bera}
\affiliation{International Centre for Radio Astronomy Research, Curtin University, Bentley, 6102, WA, Australia}

\author{C.W.~James}
\affiliation{International Centre for Radio Astronomy Research, Curtin University, Bentley, 6102, WA, Australia}
\author{J.~Paterson}
\affiliation{International Centre for Radio Astronomy Research, Curtin University, Bentley, 6102, WA, Australia}

\author{A.~T.~Deller}
\affiliation{Centre for Astrophysics and Supercomputing, Swinburne University of Technology, Hawthorn, VIC, 3122, Australia}
\author{A~C.~Gordon}
\affiliation{Center for Interdisciplinary Exploration and Research in Astrophysics (CIERA) and Department of Physics and Astronomy, Northwestern University, Evanston, IL 60208, USA}
\author{L.~Marnoch}
\affiliation{School of Mathematical and Physical Sciences, Macquarie University, NSW 2109, Australia}
\alsoaffiliation{Astrophysics and Space Technologies Research Centre, Macquarie University, Sydney, NSW 2109, Australia}
\alsoaffiliation{Australia Telescope National Facility, CSIRO Space \& Astronomy, Box 76 Epping, NSW 1710, Australia}
\author{A.~R.~Muller}
\affiliation{Maria Mitchell Observatory, Nantucket, MA 02554, USA}
\alsoaffiliation{Max Planck Institute for Gravitational Physics (Albert Einstein Institute), 14476 Potsdam, Germany}
\author{J.~Xavier~Prochaska}
\affiliation{Department of Astronomy and Astrophysics, University of California, Santa Cruz, CA 95064, USA}
\alsoaffiliation{Kavli Institute for the Physics and Mathematics of the Universe (Kavli IPMU), 5-1-5 Kashiwanoha, Kashiwa, 277-8583, Japan}
\alsoaffiliation{Division of Science, National Astronomical Observatory of Japan, 2-21-1 Osawa, Mitaka, Tokyo, 181-8588, Japan}
\author{S.~D.~Ryder}
\affiliation{School of Mathematical and Physical Sciences, Macquarie University, NSW 2109, Australia}
\alsoaffiliation{Astrophysics and Space Technologies Research Centre, Macquarie University, Sydney, NSW 2109, Australia}
\author{R.~M.~Shannon}
\affiliation{Centre for Astrophysics and Supercomputing, Swinburne University of Technology, Hawthorn, VIC, 3122, Australia}

\author{N.~Tejos}
\affiliation{Instituto de F\'isica, Pontificia Universidad Cat\'olica de Valpara\'iso, Casilla 4059, Valpara\'iso, Chile}
\author{A.~G.~Mannings}
\affiliation{Department of Astronomy and Astrophysics, University of California, Santa Cruz, CA 95064, USA}
 %% First letter not capped

\begin{document}

\begin{abstract}
Impulsive radio signals such as fast radio bursts (FRBs) are imprinted with the signatures of multi-path propagation through ionised media in the form of frequency-dependent temporal broadening of the pulse profile, commonly referred to as scattering. The dominant source of scattering for most FRBs is expected to be within their host galaxies, an assumption which can be tested by examining potential correlations between the scattering properties of the FRBs and global properties of their hosts. 
Using results from the Commensal Real-time ASKAP Fast Transient (CRAFT) survey, we investigate correlations across a range of host galaxy properties against attributes of the FRB that encode propagation effects: scattering timescale $\tau$, polarisation fractions, and absolute Faraday rotation measure. From 21 host galaxy properties considered, we find three that are correlated with $\tau$, including the stellar surface density (or compactness; Pearson p-value $p$ = 0.002 and Spearman $p$ = 0.010), the mass-weighted age (Spearman p-value $p$ = 0.009), and a weaker correlation with the gas-phase metallicity (Spearman $p = 0.017$). Weakly significant correlations are also found with $H\alpha$ equivalent widths and stellar gravitational potential. From 10,000 trials of reshuffled datasets, we expect 2 strong Spearman correlations only 2\% of the time, and three weaker correlations in 6.6\% of cases. Compact host galaxies may have more ionised content which scatters the FRB further. Compact galaxies were also found to correlate with gas-phase metallicity in our sample, while H\,{\sc ii} regions along the line-of-sight %or housing FRB sources 
are also a potential contributing factor. 
No correlation is seen with host galaxy inclination, which weakens the case for an inclination bias, as previously suggested for samples of localised FRBs. 
A strong ($p = 0.002$) correlation is found for absolute rotation measure with optical disc axis ratio b/a; greater rotation measures are seen for edge-on host galaxies. %;  suggesting that the local environment of the FRB progenitor may be more significant. 
%The fraction of circular polarisation in the FRB signal is weakly dependent on the mass-weighted age ($p = 0.028$).
%weakly dependent on the galaxy effective radius, while the linear polarisation fraction is increased for host galaxies with higher stellar metallicity. 
Further high-time resolution FRB detections, coupled with localisation and detailed follow-up on their host galaxies, are necessary to corroborate these initial findings and shed further light into the FRB mechanism.  
\end{abstract}

\section{Introduction}

Fast radio bursts (FRBs) are intense pulses of radio emission occurring on timescales of milliseconds, first discovered by \citet{Lorimer2007}. Despite the unknown nature of their progenitors, FRBs have proven to be excellent probes of the ionised gas along their line of sight \citep{Macquart2020}, as well as powerful tools for conducting cosmological studies \citep{James2022,Baptista2023,Glowacki2024c}. An important aspect in understanding the propagation of FRBs throughout the Universe en route to our radio telescopes is the effect of scattering. Due to electron density fluctuations, FRBs can propagate along multiple ray paths, leading to pulse broadening in the temporal domain, or scintillation in the frequency domain \citep{Macquart2013}. 
Understanding scattering effects can help resolve the scattering region of the source \citep{Simard2020}, and properties of the local environment \citep[e.g.][]{Ocker2022,Sammons2023}. However, scattering also acts to reduce the detectability of FRBs and smear out details of the pulse \citep[see review by][]{Cordes2019}. Furthermore, contributions to the observed scattering timescale, $\tau_{\rm obs}$, can come from the Milky Way, intervening halos, the FRB host galaxy, and/or the immediate environment of the progenitor \cite[e.g.][]{Chawla2022, Ocker2023}. These must be disentangled to properly determine the propagation path of FRBs and the contributions toward the total dispersion measure (DM) of each medium an FRB traverses through, and inform models of the likely turbulent environment near the FRB progenitor. Understanding the contribution of the host to FRB observables such as DM can also be used to inform cosmological studies possible through localised FRBs through standardisation of datasets, e.g. if the host contribution to DM is correlated, on average, with the degree of temporal broadening.

Similarly, the Faraday rotation measure (RM) of the FRB pulse, as well as the fraction of linear and circular polarisation, can be compared with the different regions FRBs travel through. For example, a property of the FRB pulse may be governed by the magnetic fields of the host galaxy rather than the local environment of the progenitor (\cite[for which magnetars are a popular model, motivated by][]{Bochenek2020}). If this were the case, then this may be reflected through correlations with global galaxy properties, e.g. depolarisation due to galaxy-level turbulence. A lack of correlation for e.g. RM with host galaxy properties would also aid in our testing of FRB progenitor models and the contribution of a more localised region to this property of the observed burst.

%While hundreds of FRBs have been published, the majority are yet to be localised to their host galaxies, with most exceptions coming from repeating FRBs which make up a minority of the total population. This is tied to the currently unknown origin of FRBs, although many theories exist \cite[see review by][]{Cordes2019}, and it is possible there is more than one progenitor to the observed population. The localisation of FRBs via their radio signals to sub-arcsecond precision is hence a necessity to help distinguish between these competing theories, and enable studies of the host galaxies \citep{Gordon2023}. 

High time resolution FRB datasets, detected at high signal-to-noise, are necessary to properly measure these FRB burst properties, including scattering timescales. % while preserving the finer details in FRB pulse structures. 
A polarimetric analysis of 128 non-repeating FRBs from the Canadian Hydrogen Intensity Mapping Experiment \citep[CHIME/FRB;][]{chime2021} was presented by \citet{Pandhi2024}, which took advantage of voltage data at a time resolution of 2.56~$\mu$s. The Commensal Real-time ASKAP Fast Transients \citep[CRAFT;][]{Macquart2010,Bannister2017} survey with the Australian Square Kilometre Array Pathfinder telescope \citep[ASKAP;][]{Deboer2009,Hotan2021} recently presented microsecond-resolution, coherently-dedispersed, polarimetric measurements of 35 FRBs \citep{Scott2025}. \citet{Sherman2024} meanwhile presented polarimetry of 25 non-repeating FRBs for the Deep Synoptic Array (DSA-110), at a time resolution of 32.768~$\mu$s. The studies with ASKAP and DSA-110 already come with associations with host galaxies, which is not the case for the majority of CHIME-detected FRBs; however, CHIME outriggers will achieve this {\myedit \citep{FRBCollaboration2025}}. With confident localisation of FRBs to (and even within) their host galaxies, one can start comparing the host galaxy properties of the FRB to their burst properties, including $\tau_{\rm obs}$, and determine if any correlations exist. This was recently investigated by \citet{Acharya2025}, with no correlation found with either stellar mass or star formation rate, albeit with a heterogeneous sample of $<20$ FRB hosts from the literature where some had only scattering time upper limits, and estimates of scattering times rather than a direct measurement from the FRB pulse. {\myedit \citet{Li2025} examined correlations of FRB host galaxy properties with excess DM, but not with other properties of the FRB pulse.}

%Another survey which enables an investigation into scattering is the Commensal Real-time ASKAP Fast Transients \citep[CRAFT;][]{Macquart2010,Bannister2017} survey with the Australian Square Kilometre Array Pathfinder telescope \citep[ASKAP;][]{Deboer2009,Hotan2021}. %CRAFT has determined FRB localisations to sub-arcsecond accuracy, by dumping 3.1~s voltage data at the time of FRB detection with corresponding bandpass and polarisation calibrator voltage data, and post-processing this voltage data. Previous localisations of CRAFT FRBs have included the highest redshift FRB host to date at $z > 1$ \citep{Ryder2023}, a dwarf host galaxy with large excess DM hosting a non-repeating FRB \citep{Bhandari2023}, and the first commensal detection of an FRB with the neutral hydrogen gas in its host galaxy \citep{Glowacki2023}. \citet{Shannon2024} presented the localisations for the incoherent sum-mode (ICS) survey, for which an in-depth analysis of the optically derived global galaxy properties of a subset of CRAFT FRBs was presented in \citet{Gordon2023}. With the high-time resolution FRB properties, including derived scattering timescale measurements, presented in \textbf{Scott et al. 2025}, analysis of FRB scattering with several host galaxy properties, including metallicity, galaxy inclination, and mass-weighted age for a sample of over 10 FRBs in each measure can be conducted. This preliminary study provides an initial framework for follow-up studies expected to follow a fast rise in high-accuracy FRB localisations over the next several years.

In this paper, we present investigations into possible correlations between FRB pulse properties (scattering timescales, linear and circular polarisation fractions, and rotation measure) and a range of host galaxy properties. Our homogeneous sample of 44 FRBs is primarily made up of CRAFT ICS-survey FRBs with high-precision localisation enabling host galaxy identification \citep{Shannon2024}, coupled with high sensitivity and high-time resolution burst profiles  \citep{Scott2025}. In Section~\ref{sec:2} we describe the sample and properties from the FRB burst and host galaxy, and the correlations examined between these aspects. In Section~\ref{sec:results} we discuss the derived correlations and potential explanations of statistically significant results, as well as an investigation into the level of significance. We summarise our findings in Section~\ref{sec:conclusion}.

%We can investigate the latter - i.e. scattering due to the circumburst environment of the FRB, which could contribute $>$ms worth of scattering time in the burst - by investigating if any correlations exist between scattering times and FRB host galaxy properties. 

\section{Sample and methods}\label{sec:2}

\subsection{FRB sample and methods}\label{sec:frbsample}

%subsubsection{Scattering properties properties}

We begin with the CRAFT incoherent sum mode (ICS) sample as presented in \citet{Scott2025} of 35 FRBs and also included in \citet{Shannon2024}. The high-time resolution datasets were derived through the CRAFT Effortless Localisation and Enhanced Burst Inspection pipeline \citep[CELEBI;][]{Scott2023}. CELEBI is an automated offline software pipeline that extends previous software of the CRAFT survey team to process ASKAP voltages in order to produce sub-arcsecond precision localisations of FRB events, alongside polarimetric data at 3~ns time resolution. Following flagging and calibration of the 3.1~second voltage data, an image is made using the time gated period around the FRB event, in order to isolate the FRB emission entirely and maximise the signal to noise (S/N) in the FRB image \citep[Fig{\myedit ure}\ 5 of][]{Scott2023}. The downloaded ASKAP voltages are then beamformed on the derived FRB position, producing high-time resolution datasets with full polarisation information.

In addition, we consider eight other localised FRBs with pulse properties (Section~\ref{sec:frbprop}), %host galaxies with stellar masses and star formation rate (SFR) information, 
seven of which have reliable scattering timescale measurements in the literature \citep{CHIMEFRBCollaboration2021,Rajwade2022,Driessen2024,Connor2023,Cassanelli2024}. While there are more FRBs not already covered by the CRAFT ICS sample in the study conducted by \citet{Acharya2025}, we have excluded FRBs with only upper limits present, or scattering time values presented in other literature without errors or stated to be estimates based on scintillation rather than a direct measurement of $\tau$. 

FRB\,20190520B is excluded due to its sightline intersecting multiple foreground galaxy clusters \citep{Lee2023}, which may lead to the observed scattering being dominated by effects produced outside of the FRB host galaxy. We note that FRB\,20190714A, and to a lesser extent FRB\,20200906A, were also found to have foreground galaxies near their FRB sightlines contributing towards excess DM (DM attributed to the foreground galaxies rather than the FRB host) in a study by \citet{Simha2023}, albeit they ``do not find any group contribution when applying our
fiducial halo gas model, which truncates at the virial radius, to the groups identified in this field". Excluding one or both of these FRBs did not significantly alter results presented in Section~\ref{sec:results}.
%Table~\ref{tab:burstprop} presents the scattering times, RMs, and polarisation fractions and global host galaxy measurements considered in this work. 

\begin{table*}[h]
\centering
\caption{FRB pulse properties, where we list the FRB name; host galaxy redshift; logarithm of the rest-frame scattering time at 1~GHz; {\myedit $\alpha$; whether a single component (s) or multiple (m) components were used to fit for scattering $N_{\tau}$ (where available - `a' indicates ambiguity - see section~2.3.7 of \citet{Scott2025} for further details);} logarithm of the absolute RM; linear polarisation fraction; circular polarisation fraction; and total polarisation fraction. CRAFT FRBs with potentially unreliable polarisation fraction measures have their values indicated with *, and these values are not used in the main analysis. With the exception of the last eight FRBs in the table, these FRB burst properties are derived by \citet{Scott2025}. References for the FRBs from the literature: {a: \citet{CHIMEFRBCollaboration2021}, b: \citet{Rajwade2022}, c: \citet{Driessen2024}, d: \citet{Connor2023}, e: \citet{Cassanelli2024}, f: \citet{Caleb2023}}.}\label{tab:burstprop}
\addtolength{\tabcolsep}{-0.35em}
\begin{tabular}{llcccccccccccc}
\toprule
    FRB &   $z$ & log($\tau_{\rm 1 GHz}$x(1+$z$)$^{3}$)& Error & $\alpha$ & $N_{\tau}$ & log(|RM$_{\rm ex}$|x(1+$z$)$^{2}$) & Error & Pol$_{\rm lin}$ & Error & Pol$_{\rm circ}$ &Error & Pol$_{\rm tot}$ & Error \\
     &    & ms & ms   &  & & rad m$^{-2}$ & rad m$^{-2}$ &  &  &  & & &  \\
\midrule
 FRB20180916B &    0.0337 &  &  &  &  & 1.235 &  0.013 &  -- &   -- &    -- & -- & -- &  -- \\
 FRB20180924B &    0.3212 &  &  &  &  & 0.418 &  0.015 & 0.89 & 0.02 &  0.09 &   0.02 & 0.9 & 0.02 \\
 FRB20181112A &    0.4755 & --0.902 & 0.021 & --2.0$\pm$0.3 &  m  & 1.093 &  0.054 & 0.92 &    0 &  0.19 & 0 &    0.94 &   0 \\
 FRB20190102C &    0.2912 & --0.713 &  0.134 & --5.5$\pm$1 & m & 2.346 &  0.120 & 0.86 & 0.01 &   0.1 &   0.01 &    0.86 & 0.01 \\
 FRB20190608B &  0.1178 & 1.074 &  0.044 & --3.37$\pm$1.3 & s & 2.671 &  0.083 &   s & 0.04 &  0.02 &   0.02 &  1 & 0.04 \\
 FRB20190611B &    0.3778 &  &   &  &  & 1.183 &  0.035 & 0.75 & 0.05 &  0.29 &   0.04 & 0.8 & 0.04 \\
 FRB20190711A &   0.5217 & --1.404 &  0.055 & --2.5$\pm$1.1 & m & 1.553 &  0.079 &  0.98 &   0.03 &    0.14 & 0.03 & 0.99 &  0.03 \\
 FRB20190714A &    0.2365 & 0.196 &  0.014 & --2.7$\pm$0.6 &  s  & -- & -- &  -- &   -- &    -- & -- & -- &  -- \\
 FRB20191001A & 0.234 & 0.524 &  0.005 & --4.85$\pm$0.3 & s & 1.615 &  0.144 & 0.53 & 0.01 &  0.05 &   0.01 &    0.54 & 0.01 \\
 FRB20191228B &    0.2432 & 1.430 &  0.016  & --3.6$\pm$0.6 & s & 1.028 &  0.059 & 0.92 & 0.02 &  0.15 &   0.02 &    0.93 & 0.02 \\
 FRB20200430A & 0.161 & 0.915 &  0.013 & --1.45$\pm$0.2 & s & 2.387 &  0.140 &  0.43 & 0.02 &  0.04 & 0.02 & 0.43 & 0.02 \\
 FRB20200906A &    0.3688 & --1.421 &  0.005 & --4.5$\pm$0.4 & s &  &   &  0.8 & 0.005 &  0.073 & 0.004 & 0.804 & 0.005 \\
 FRB20210117A &    0.2145 &  &  &  &   & 1.856 &  0.079 & 0.93 & 0.02 &  0.05 &   0.01 &    0.92 & 0.02 \\
 FRB20210320C &    0.2797 & --0.754 &  0.005 & --4.4$\pm$0.1 & m & 2.679 &  0.189 &  *0.86 & *0.008 &    *0.117 & *0.006 & *0.868 &  *0.008 \\
 FRB20210807D &   0.1293 &  &    &  & & 1.682 &  0.030 &  -- &   -- &    -- & -- & -- &  -- \\
 FRB20211127I &    0.0469 &  &    &  & & 1.850 &  0.126 &  0.244 &  0.003 & 0.129 & 0.003 & 0.276 & 0.003 \\
 FRB20211203C &   0.3439 & 0.125 &  0.033 & --9.7$\pm$2.4 & s & 2.057 &  0.085 &  0.57 & 0.02 & 0.07 & 0.03 & 0.58 & 0.02 \\
 FRB20211212A &    0.0707 & 0.992 &  0.265 & --2.8$\pm$2.3 & s & 1.741 &  0.105 & *0.47 & *0.02 &  *0.09 &   *0.02 &    *0.48 & *0.02 \\
 FRB20220105A &    0.2785 & 0.399 &  0.087 &  --2$\pm$0.8 & m  &  3.323 &  0.112 & 0.3 & 0.03 &  0.05 &   0.03 & 0.3 & 0.03 \\
 FRB20220501C & 0.381 &  &   &  &  & 1.688 &  0.135 & 0.68 & 0.02 &  0.06 &   0.02 &    0.69 & 0.02 \\
 FRB20220610A & 1.015 & 0.845 &  0.000 & --3.56$\pm$0.03 & s & 2.920 &  0.190 & 0.98 & 0.01 &  0.06 &  0.007 &    0.98 & 0.01 \\
 FRB20220725A &    0.1926 & 0.520 &  0.007 & 1.94$\pm$0.06 & a & 2.370 &  0.019 & *0.58 & *0.02 &  *0.13 &   *0.03 & *0.6 & *0.02 \\
 FRB20220918A &    0.4908 & 1.520 &  0.014 & --2.10$\pm$0.03 & s & 3.022 &  0.042 & 0.15 & 0.01 &  0.11 &   0.02 &    0.19 & 0.02 \\
 FRB20221106A & 0.204 &  &   &  &  & 2.774 &  0.099 &    0.862 & 0.008 & 0.078 &  0.006 &   0.865 &    0.008 \\
 FRB20230526A & 0.157 & 0.629 &  0.010 & --3.6$\pm$0.3 & a & 2.907 &  0.187 &  0.39 &  0.008 & 0.04 & 0.008 & 0.393 & 0.008 \\
 FRB20230708A & 0.105 & --0.548 &  0.008 & --2.84$\pm$0.4 & m & 1.794 &  0.069 & 0.95 & 0.01 &  0.39 &  0.008 &   1.031 &    0.009 \\
 FRB20230718A & 0.035 & --0.725 &  0.046 & --1.6$\pm$0.4 & m & 1.783 &  0.015 & 0.92 & 0.02 &  0.11 &   0.01 &    0.92 & 0.02 \\
 FRB20230902A &    0.3619 & --0.740 &  0.005 & --2.55$\pm$0.08 & m & 2.458 &  0.151 & 0.91 & 0.01 &  0.05 &   0.01 &    0.91 & 0.01 \\
 FRB20231226A &    0.1539 &  &   &  &  & 2.743 &  0.168 & 0.86 & 0.02 &  0.04 &   0.01 &    0.86 & 0.02 \\
 FRB20240201A &  0.042729 & --0.283 &  0.050 & --3.9$\pm$0.5 & m & 3.140 &  0.208 &  *0.76 & *0.02 &    *0.09 & *0.02 & *0.76 & *0.02 \\
 FRB20240208A &    0.2385 & 0.279 &  0.103 & --2.7$\pm$2.1 & s & 2.066 &  0.110 &  0.94 &   0.09 &    0.08 & 0.08 & 0.94 &  0.09 \\
 FRB20240210A &  0.023686 & --0.199 &  0.027 & --3.6$\pm$0.3 & m & 2.533 &  0.307 &  *0.73 &   *0.02 &    *0.14 & *0.02 & *0.74 &  *0.02 \\
 FRB20240304A &    0.2423 & 0.885 &  0.028 & 3.5$\pm$1.3 & s &  &   &  0.92 &   0.03 &    0.04 & 0.02 & 0.92 &  0.03 \\
 FRB20240310A & 0.127 & 0.270 &  0.187 & --3.23$\pm$0.5 & s & 3.336 &  0.189 & 0.75 & 0.03 &  0.05 &   0.03 &    0.75 & 0.03 \\
 FRB20240318A & 0.112 & --0.754 &  0.005 & --3.32$\pm$0.005 & m & 2.791 &  0.498 &  0.8 &   0.02 &    0.13 & 0.01 & 0.81 &  0.02 \\
  \hline
 FRB20181030A$^{a}$ &    0.0039 &  &  &   &  & 1.751 &  0.110 &  -- &   -- &    -- & -- & -- &  -- \\
 FRB20181220A$^{a}$ & 0.027 & --1.194 &  0.027 & &   &  &   &  -- &   -- &    -- & -- & -- &  -- \\
 FRB20181223C$^{a}$ & 0.03 & --1.848 &  0.092 &  &   & &   &  -- &   -- &    -- & -- & -- &  -- \\
 FRB20210410D$^{f}$ &    0.1415 & 1.641 &  0.028 & --4 & s  & 1.705 &  0.058 &  -- &   -- &    -- & -- & -- &  -- \\
 FRB20201123A$^{b}$ &    0.0507 & 0.939 &  0.000 & --4.2$\pm$0.4 & s &   &   &  -- &   -- &    -- & -- & -- &  -- \\
 FRB20210405I$^{c}$ & 0.066 & 1.070 &  0.007 & --4.6$\pm$0.1 &  s &  &   &  -- &   -- &    -- & -- & -- &  -- \\
 FRB20220509G$^{d}$ &    0.0894 & --0.411 &  0.078 & &   & 2.107 &  0.060 &  -- &   -- &    -- & -- & -- &  -- \\
 FRB20190110C$^{a}$ &    0.1224 & --1.402 &  0.044 & &   & 2.019 &  0.392 &  -- &   -- &    -- & -- & -- &  -- \\
 FRB20210603A$^{e}$ & 0.177 & --1.487 &  0.005 & &   & 2.294 &  3.318 &  -- &   -- &    -- & -- & -- &  -- \\
\bottomrule
\end{tabular}
\end{table*}

\afterpage{
    \clearpage% Flush earlier floats (otherwise order might not be correct)
    \pagestyle{empty}% empty page style (?)
    \newgeometry{left=0.5cm,right=0.5cm,bottom=0.5cm,top=0.5cm} 
    \begin{landscape}% Landscape page
        \begin{table}[h!]
        \setlength\extrarowheight{-0.25pt}
        \addtolength{\tabcolsep}{-0.40em}
        %\centering % Center table
        \begin{tabular}{llllllllllllllllllllllllllll}
\toprule
 FRB & z & SF? & $m_{R}$ & log($M_{\rm F}$) & Error & log($M_{\rm *}$) & Error & log(SFR) & Error & log($\frac{SFR}{M}$) & Error & $F_{\rm H}$\textsubscript{$\alpha$} 10$^{-16}$ & Error 10$^{-16}$ & $H\alpha$ EW & Error & $A_{\rm V,o}$ & Error & $A_{\rm V,y}$ & Error & $\frac{Z_{\rm gas}}{Z_{\odot}}$ & Error & $\frac{Z_{\rm star}}{Z_{\odot}}$ & Error & $t_{\rm m}$ & Error  & [S\,{\sc ii}] ratio & Error \\
   &  &  & mag & M$_{\odot}$ & M$_{\odot}$ & M$_{\odot}$ & M$_{\odot}$ & M$_{\odot}$ yr$^{-1}$ & M$_{\odot}$ yr$^{-1}$ & yr$^{-1}$ & yr$^{-1}$ & {\scriptsize erg\,s$^{-1}$\,cm$^{-2}$} & {\scriptsize erg\,s$^{-1}$\,cm$^{-2}$} & \AA & \AA & mag & mag & mag & mag &   &   &   &   & Gyr  & Gyr & &  \\
 %[1e-16 erg/s/cm\textasciicircum 2]
\midrule
 FRB20180916B & 0.0337 & N & 16.17 & 10.13 & 0.045 & 9.91 & 0.04 & -1.40 & 0.27 & -11.31 & 0.27 & 96.75 & 2.68 & 6.79 & 0.39 & 0.35 & 0.04 & 0.94 & 0.26 & 1.51 & 0.63 & 0.02 & 0.00 & 7.73 & 1.04 & & \\
 FRB20180924B & 0.3212 & Y & 20.33 & 10.6 & 0.025 & 10.39 & 0.02 & -0.21 & 0.20 & -10.60 & 0.20 & 3.67 & 0.08 & 36.95 & 1.91 & 0.11 & 0.03 & 1.1 & 0.28 & 1.07 & 0.20 & 0.72 & 0.07 & 5.63 & 0.64 & &\\
 FRB20181112A & 0.4755 & Y & 21.68 & 10.06 & 0.075 & 9.87 & 0.07 & 0.19 & 0.23 & -9.68 & 0.24 & 1.76 & 0.07 & 14.08 & 0.98 & 0.13 & 0.10 & 1.16 & 0.28 & 0.68 & 0.19 & 0.65 & 0.44 & 3.82 & 0.91 & & \\
 FRB20190102C & 0.2912 & Y & 20.77 & 9.9 & 0.09 & 9.69 & 0.10 & -0.40 & 0.23 & -10.09 & 0.25 & 3.78 & 0.29 & 25.79 & 2.75 & 0.2 & 0.15 & 1.09 & 0.29 & 0.31 & 0.46 & 0.07 & 0.06 & 4.76 & 1.25 & & \\
 FRB20190608B & 0.1178 & Y & 17.41 & 10.78 & 0.02 & 10.56 & 0.02 & 0.85 & 0.02 & -9.71 & 0.03 & 74.48 & 1.63 & 17.44 & 0.90 & 0.08 & 0.02 & 1.09 & 0.21 & 1.05 & 0.12 & 0.93 & 0.09 & 7.13 & 0.95 & & \\
 FRB20190611B & 0.3778 & Y & 22.15 & 9.77 & 0.13 & 9.57 & 0.12 & -0.28 & 0.42 & -9.85 & 0.44 & 1.48 & 0.1 & 31.32 & 3.06 & 0.45 & 0.29 & 1.2 & 0.29 & 1.00 & 0.61 & 0.14 & 0.18 & 4.45 & 1.16 & & \\
 FRB20190711A & 0.5217 & Y & 23.54 & 9.29 & 0.21 & 9.10 & 0.19 & -0.02 & 0.33 & -9.12 & 0.38 & & & & & 0.28 & 0.20 & 1.06 & 0.27 & & & 0.10 & 0.13 & 3.54 & 1.16 & & \\
 FRB20190714A & 0.2365 & Y & 20.34 & 10.42 & 0.045 & 10.22 & 0.04 & 0.28 & 0.11 & -9.94 & 0.12 & 9.61 & 0.24 & 33.24 & 1.83 & 0.69 & 0.20 & 1.05 & 0.28 & 1.32 & 0.65 & 0.81 & 0.72 & 5.48 & 0.89  & & \\
 FRB20191001A & 0.234 & Y & 18.36 & 10.92 & 0.085 & 10.73 & 0.07 & 1.26 & 0.31 & -9.47 & 0.32 & 45.4 & 0.94 & 20.8 & 1.05 & 1.06 & 0.10 & 1.15 & 0.27 & 0.83 & 0.21 & 0.30 & 0.07 & 3.89 & 1.62 & & \\
 FRB20191228B & 0.2432 & Y & & & & & & & && & & & & & & & & & & & & & & & & \\
 FRB20200430A & 0.161 & Y & 21.05 & 9.51 & 0.085 & 9.30 & 0.09 & -0.96 & 0.20 & -10.26 & 0.21 & 3.36 & 0.14 & 22.12 & 1.59 & 0.38 & 0.14 & 1.08 & 0.33 & 0.76 & 0.10 & 0.10 & 0.08 & 5.99 & 1.14 & & \\
 FRB20200906A & 0.3688 & Y & 19.95 & 10.57 & 0.055 & 10.37 & 0.05 & 0.69 & 0.26 & -9.68 & 0.26 & 14.56 & 0.38 & 51.48 & 2.89 & 0.2 & 0.10 & 1.09 & 0.25 & 0.55 & 0.17 & 0.41 & 0.15 & 4.3 & 0.98 & & \\
 FRB20210117A & 0.2145 & Y & 22.97 & 8.8 & 0.06 & 8.59 & 0.06 & -1.70 & 0.22 & -10.29 & 0.22 & 0.39 & 0.02 & 19.93 & 1.62 & 0.05 & 0.04 & 1.19 & 0.29 & 0.50 & 0.09 & 0.02 & 0.01 & 5.01 & 1.05 & & \\
 FRB20210320C & 0.2797 & Y & 19.47 & 10.57 & 0.06 & 10.37 & 0.06 & 0.55 & 0.24 & -9.82 & 0.25 & 16.11 & 0.45 & 36.08 & 2.09 & 0.64 & 0.16 & 1.26 & 0.26 & 1.02 & 0.33 & 0.15 & 0.06 & 4.56 & 1.07 & & \\
 FRB20210410D & 0.1415 & N & 20.65 & 9.7 & 0.055 & 9.47 & 0.05 & -1.52 & 0.29 & -10.99 & 0.29 & 2.81 & 0.11 & 15.7 & 1.09 & 0.39 & 0.12 & 1.14 & 0.29 & 1.07 & 0.60 & 0.09 & 0.05 & 6.78 & 1.25 & & \\
 FRB20210807D & 0.1293 & N & 17.17 & 11.2 & 0.02 & 10.97 & 0.02 & -0.20 & 0.12 & -11.17 & 0.12 & 22.89 & 0.55 & 2.97 & 0.16 & 0.04 & 0.03 & 1.08 & 0.16 & 0.55 & 0.09 & 0.30 & 0.03 & 8.36 & 2.04 & & \\
 FRB20211127I & 0.0469 & Y & 14.97 & 9.58 & 0.05 & 9.48 & 0.04 & 1.55 & 0.02 & -7.93 & 0.04 & 501.43 & 12.26 & 15.14 & 0.82 & 0.06 & 0.01 & 1.22 & 0.28 & 1.95 & 0.56 & 0.30 & 0.02 & 3.85 & 2.89 & & \\
 FRB20211203C & 0.3439 & Y & 19.64 & 9.9 & 0.095 & 9.76 & 0.08 & 1.20 & 0.08 & -8.56 & 0.11 & 11.32 & 0.29 & 42.13 & 2.34 & 0.04 & 0.03 & 1.08 & 0.26 & 0.56 & 0.19 & 1.00 & 0.36 & 2.47 & 1.62 & &  \\
 FRB20211212A & 0.0707 & Y & 16.44 & 10.49 & 0.065 & 10.28 & 0.06 & -0.14 & 0.30 & -10.42 & 0.31 & 99.33 & 4.44 & 10.89 & 0.81 & 0.19 & 0.04 & 1.19 & 0.27 & 1.58 & 0.78 & 0.17 & 0.04 & 5.83 & 1.10 & & \\
 FRB20220105A & 0.2785 & Y & 21.19 & 10.22 & 0.065 & 10.01 & 0.06 & -0.93 & 0.01 & -10.94 & 0.01 & 2.36 & 0.09 & 22.73 & 1.55 & 0.76 & 0.16 & 1.15 & 0.27 & 0.72 & 0.22 & 0.15 & 0.05 & 5.67 & 0.98 & 1.51 & 0.21 \\
 FRB20220501C & 0.381 & Y & & & & & & & & & & & & & & & & & & & & & & & & & \\
 FRB20220610A$^{a}$ & 1.015 & Y & & 10.11 & 0.18 & 9.69 & 0.11 & 0.22 & 0.43 & --9.47 & 0.44 & & & & & & & & & & & & & 2.60 & 0.91 & & \\
 FRB20220725A & 0.1926 & Y & 17.81 & & & & & 0.22 & 0.43 & & & & & & & & & & & & & & & & & 1.20 & 0.04 \\
 FRB20220918A & 0.4908 & Y & 23.58 & & & & & 0.00 & 0.00 & & & & & & & & & & & & & & & & & & \\
 FRB20221106A & 0.204 & Y & 18.32 & & & & & -0.87 & 0.01 & & & & & & & & & & & & & & & & & 1.30 & 0.11 \\
 FRB20230526A & 0.157 & Y & 21.03 & & & & & -0.98 & 0.00 & & & & & & & & & & & & & & & & & 1.45 & 0.09 \\
 FRB20230708A & 0.105 & Y & 22.53 & & & 7.97 & 0.09 & -1.48 &  & -9.62 & 0.27 & & & & & & & & & & & & & 5.82 & 1.25 & & \\
 FRB20230718A & 0.035 & Y & & & & & & & & & & & & & & & & & & & & & & & & & \\
 FRB20230902A & 0.3619 & Y & 21.49 & & & & & -0.61 & 0.01 & & & & & & & & & & & & & & & & & 1.55 & 0.19 \\
 FRB20231226A & 0.1539 & Y & 18.94 & & & & -1.05 & 0.00 & & & & & & & & & & & & & & & & & & & \\
 FRB20240201A$^{b}$ & 0.0427 & Y & 16.91 & & & 10.21 & 0.02 & 0.14 & 0.02 & -10.07 & 0.04 & & & & & & & & & & & & & & & 1.43 & 0.11 \\
 FRB20240208A & 0.2385 & Y & & & & & & & & & & & & & & & & & & & & & & & & & \\
 FRB20240210A & 0.0237 & Y & & & & & & & & & & & & & & & & & & & & & & & & & \\
 FRB20240304A & 0.2423 & Y & & & & & & & & & & & & & & & & & & & & & & & & & \\
 FRB20240310A & 0.127 & Y & & & & & & & & & & & & & & & & & & & & & & & & & \\
 FRB20240318A & 0.112 & Y & & & & & & & & & & & & & & & & & & & & & & & & & \\
 \hline 
 FRB20181030A$^{c}$ & 0.0039 & Y & & & & 9.76 & 0.00 & -0.46 & 0.00 & -10.22 & 0.00 & & & & & & & & & & & & & & & & \\
 FRB20181220A$^{c}$ & 0.027 & Y & & & & 9.86 & 0.14 & 0.46 & 0.24 & -9.40 & 0.38 & & & & & & & & & & & & & & & & \\
 FRB20181223C$^{c}$ & 0.03 & Y & & & & 9.29 & 0.20 & -0.82 & 0.35 & -10.11 & 0.55 & & & & & & & & & & & & & & & & \\
 FRB20201123A$^{d}$ & 0.0507 & Y & & & & 11.20 & 0.00 & -0.70 & 0.00 & -11.90 & 0.00 & & & & & & & & & & & & & & & & \\
 FRB20210405I$^{e}$ & 0.066 & Y & & & & 11.25 & 0.00 & -0.52 & 0.00 & -11.77 & 0.00 & & & & & & & & & & & & & & & & \\
 FRB20220509G$^{f}$ & 0.0894 & Y & & & & 10.70 & 0.01 & -0.60 & 0.12 & -11.30 & 0.13 & & & & & & & & & & & & & & & & \\
 FRB20190110C$^{c}$ & 0.1224 & N & & & & 10.75 & 0.00 & -0.23 & 0.00 & -10.98 & 0.00 & & & & & & & & & & & & & & & & \\
 FRB20210603A$^{g}$ & 0.177 & Y & & & & 10.93 & 0.04 & -0.62 & 0.11 & -11.55 & 0.15 & & & & & & & & & & & & & & & & \\
\bottomrule
\end{tabular}
        \captionof{table}{Global galaxy properties of FRB host galaxies. We list the FRB name, redshift, whether the host galaxy has been identified as star forming (as opposed to transitioning or quiescent), $R$-band AB magnitude, total stellar mass formed over the life of the galaxy, current-day stellar mass, integrated 0--100~Myr star formation rate SFR, integrated 0--100~Myr specific SFR (SFR/$M_{*}$, or sSFR), $H\alpha$ flux, %the stellar continuum at the $H\alpha$ flux, 
        the $H\alpha$ equivalent width (EW), dust extinction due to old and young stellar populations, the gas-phase metallicity, the stellar metallicity, the mass-weighted age, and the [S\,{\sc ii}] close doublet ratio. These values and associated errors (except for $H\alpha$ EW and stellar continuum) are from \citet{Gordon2023} for most sources. All [S\,{\sc ii}] close doublet ratios, some SFR, magnitudes, and $H\alpha$ fluxes, and the stellar mass and mass-weighted age measure for FRB\,20230708A are from work by \citet{Muller2025}. Other properties come from the literature for FRBs denoted with the following superscripts: {a: \citet{Ryder2023} and \citet{Gordon2024}, b: \citet{Chang2015}, c:  \citet{CHIMEFRBCollaboration2021}, d: \citet{Rajwade2022}, e: \citet{Driessen2024}, f: \citet{Connor2023}, g: \citet{Cassanelli2024}}.}\label{tab:galprop}% Add 'table' caption
        \end{table}
    \end{landscape}
    
    \clearpage% Flush page
    \restoregeometry
}

\FloatBarrier
\begin{table*}[h!]
\addtolength{\tabcolsep}{-0.40em}
\centering
\caption{Further host galaxy properties derived through {\sc galfit} for CRAFT ICS FRBs. We give the FRB name, redshift, whether the host is star-forming, the host effective radius, projected galactrocentric offset (distance of the FRB localisation from the optical centre of the host galaxy), the projected offset divided by the effective radius, the inclination of the optical disc, the optical disc semi-minor/semi-major axis ratio b/a, the logarithm of the potential log($M_{*}$/$R_{\rm eff}$), and logarithm of the compactness log($M_{*}$/($R_{\rm eff}$)$^{2}$). All values besides potential and compactness are presented in Marnoch et al., in prep.}\label{tab:galfitprop} 
\begin{tabular}{lllllllllllllllllllll}
\toprule
  &  & & $R_{\rm eff}$ & Error & log($R_{\rm eff}$) & Error & Offset & Error & $\frac{Offset}{R_{\rm eff}}$ & Error & Incl. & Error & b/a & Error & log($\frac{M_{*}}{R_{\rm eff}}$) & Error & log($\frac{M_{*}}{R_{\rm eff}^{2}}$) & Error  \\
  FRB & z & SF? & kpc & kpc & kpc & kpc & kpc & kpc &  &  & deg & deg &  &  & {\tiny M$_{\odot}$ kpc$^{-1}$} & {\tiny M$_{\odot}$ kpc$^{-1}$} & {\tiny M$_{\odot}$ kpc$^{-2}$} & {\tiny M$_{\odot}$ kpc$^{-2}$} \\
\midrule
 FRB20180916B & 0.0337 & N & & & & & & & & & & & & & & & & \\
 FRB20180924B & 0.3212 & Y & 7.74 & 0.34 & 0.889 & 0.019 & 3.86 & 0.94 & 0.50 & 0.29 & 49.34 & 0.66 & 0.669 & 0.008 & 9.50 & 0.04 & 8.61 & 0.08 \\
 FRB20181112A & 0.4755 & Y & 5.51 & 0.23 & 0.741 & 0.018 & 2.37 & 13.64 & 0.43 & 5.80 & 27.41 & 3.10 & 0.89 & 0.02 & 9.13 & 0.09 & 8.39 & 0.18 \\
 FRB20190102C & 0.2912 & Y & 4.71 & 0.06 & 0.673 & 0.006 & 1.35 & 2.67 & 0.29 & 2.00 & 62.78 & 1.04 & 0.49 & 0.01 & 9.02 & 0.11 & 8.34 & 0.21 \\
 FRB20190608B & 0.1178 & Y & 4.33 & 0.05 & 0.636 & 0.005 & 6.38 & 0.88 & 1.47 & 0.15 & 59.21 & 0.17 & 0.54 & 0.002 & 9.92 & 0.02 & 9.29 & 0.05\\
 FRB20190611B & 0.3778 & Y & 2.57 & 0.13 & 0.410 & 0.021 & 11.40 & 6.94 & 4.44 & 0.66 & 39.48 & 5.48 & 0.78 & 0.06 & 9.16 & 0.14 & 8.75 & 0.28 \\
 FRB20190711A & 0.5217 & Y & 3.44 & 2.72 & 0.536 & 0.344 & 8.75 & 9.65 & 2.55 & 1.89 & 61.57 & 12.62 & 0.5 & 0.2 & 8.56 & 0.53 & 8.03 & 1.07 \\
 FRB20190714A & 0.2365 & Y & 4.02 & 0.01 & 0.605 & 0.001 & 2.91 & 1.67 & 0.72 & 0.58 & 72.00 & 0.12 & 0.363 & 0.002 & 9.62 & 0.04 & 9.01 & 0.08 \\
 FRB20191001A & 0.234 & Y & 6.25 & 0.02 & 0.796 & 0.001 & 3.93 & 1.57 & 0.63 & 0.40 & 62.60 & 0.12 & 0.493 & 0.002 & 9.93 & 0.08 & 9.14 & 0.15 \\
 FRB20191228B & 0.2432 & Y & 2.17 & 0.06 & 0.337 & 0.013 & 4.38 & 2.72 & 2.02 & 0.65 & 60.27 & 1.86 & 0.53 & 0.02 & & & & \\
 FRB20200430A & 0.161 & Y & 1.99 & 0.01 & 0.299 & 0.003 & 2.89 & 1.20 & 1.45 & 0.42 & 75.56 & 0.45 & 0.316 & 0.006 & 9.00 & 0.09 & 8.70 & 0.18 \\
 FRB20200906A & 0.3688 & Y & 9.01 & 0.02 & 0.955 & 0.001 & 6.45 & 2.45 & 0.72 & 0.38 & 78.99 & 0.13 & 0.274 & 0.002 & 9.42 & 0.05 & 8.46 & 0.10 \\
 FRB20210117A & 0.2145 & Y & 2.84 & 1.49 & 0.454 & 0.228 & 3.53 & 1.47 & 1.24 & 0.94 & 61.94 & 6.18 & 0.5 & 0.08 & 8.14 & 0.28 & 7.68 & 0.57 \\
 FRB20210320C & 0.2797 & Y & 4.87 & 0.02 & 0.688 & 0.002 & 2.22 & 1.61 & 0.46 & 0.73 & 48.13 & 0.23 & 0.684 & 0.003 & 9.68 & 0.06 & 8.99 & 0.11 \\
 FRB20210410D & 0.1415 & N & & & & & & & & & & & & & & & & \\
 FRB20210807D & 0.1293 & N & 15.13 & 0.20 & 1.180 & 0.006 & 8.92 & 0.96 & 0.59 & 0.12 & 46.17 & 0.32 & 0.707 & 0.004 & 9.79 & 0.03 & 8.61 & 0.05 \\
 FRB20211127I & 0.0469 & Y & 10.11 & 0.05 & 1.005 & 0.002 & 2.10 & 0.50 & 0.21 & 0.24 & 18.03 & 0.31 & 0.953 & 0.003 & 8.48 & 0.04 & 7.47 & 0.08 \\
 FRB20211203C & 0.3439 & Y & 2.41 & 0.01 & 0.383 & 0.001 & 3.01 & 1.85 & 1.25 & 0.62 & 25.72 & 0.43 & 0.905 & 0.003 & 9.38 & 0.08 & 8.99 & 0.16 \\
 FRB20211212A & 0.0707 & Y & 4.41 & 0.01 & 0.644 & 0.001 & 2.16 & 0.68 & 0.49 & 0.32 & 35.58 & 0.16 & 0.822 & 0.002 & 9.64 & 0.06 & 8.99 & 0.11 \\
 FRB20220105A & 0.2785 & Y & 3.01 & 0.03 & 0.478 & 0.005 & 7.91 & 3.87 & 2.63 & 0.50 & & & 0259 & 0.001 & 9.53 & 0.06 & 9.05 & 0.13 \\
 FRB20220501C & 0.381 & Y & 2.50 & 0.10 & 0.398 & 0.017 & 2.64 & 1.84 & 1.06 & 0.74 & 36.65 & 4.83 & 0.81 & 0.05 & & & & \\
 FRB20220610A & 1.015 & Y & & & & & & & & & & & & & & & & \\
 FRB20220725A & 0.1926 & Y & 5.80 & 0.03 & 0.763 & 0.002 & 1.53 & 1.18 & 0.26 & 0.78 & 58.58 & 0.17 & 0.548 & 0.002 & & & & \\
 FRB20220918A & 0.4908 & Y & 1.71 & 0.07 & 0.232 & 0.019 & 2.89 & 2.22 & 1.69 & 0.81 & 53.35 & 4.40 & 0.62 & 0.06 & & & & \\
 FRB20221106A & 0.204 & Y & 6.72 & 0.05 & 0.827 & 0.003 & 3.90 & 1.93 & 0.58 & 0.50 & 50.90 & 0.16 & 0.65 & 0.002 & & & & \\
 FRB20230526A & 0.157 & Y & & & & & & & & & & & & & & & & \\
 FRB20230708A & 0.105 & Y & 0.58 & 0.08 & -0.238 & 0.061 & 0.33 & 0.63 & 0.57 & 2.05 & 66.19 & 4.50 & 0.44 & 0.06 & 8.21 & 0.06 & 8.45 & 0.12 \\
 FRB20230718A & 0.035 & Y & & & & & & & & & & & & & & & & \\
 FRB20230902A & 0.3619 & Y & 2.14 & 0.02 & 0.330 & 0.004 & 2.63 & 2.38 & 1.23 & 0.92 & 69.67 & 0.77 & 0.39 & 0.01 & & & & \\
 FRB20231226A & 0.1539 & Y & 5.83 & 0.04 & 0.765 & 0.003 & 5.91 & 1.41 & 1.01 & 0.25 & 44.64 & 0.22 & 0.725 & 0.003 & & & & \\
 FRB20240201A & 0.0427 & Y & 4.13 & 0.01 & 0.616 & 0.001 & 9.63 & 3.78 & 2.33 & 0.39 & & & 0.166 & 0.0002 & 9.59 & 0.02 & 8.98 & 0.18 \\
 FRB20240208A & 0.2385 & Y & & & & & & & & & & & & & & & & \\
 FRB20240210A & 0.0237 & Y & 3.90 & 0.01 & 0.591 & 0.001 & 2.05 & 0.41 & 0.53 & 0.20 & 54.35 & 0.04 & 0.6051 & 0.0005 & & & & \\
 FRB20240304A & 0.2423 & Y & 4.13 & 0.04 & 0.616 & 0.004 & 0.86 & 1.57 & 0.21 & 1.84 & 51.77 & 0.42 & 0.638 & 0.005 & & & & \\
 FRB20240310A & 0.127 & Y & 3.21 & 0.01 & 0.506 & 0.001 & 4.64 & 0.26 & 1.45 & 0.06 & 84.58 & 0.27 & 0.22 & 0.002 & & & & \\
 FRB20240318A & 0.112 & Y & & & & & & & & & & & & & & & & \\
% \hline 
% FRB20181030A & 0.0039 & Y & & & & & & & & & & & \\
% FRB20181220A & 0.027 & Y & & & & & & & & & & & \\
% FRB20181223C & 0.03 & Y & & & & & & & & & & & \\
% FRB20201123A & 0.0507 & Y & & & & & & & & & & & \\
% FRB20210405I & 0.066 & Y & & & & & & & & & & & \\
% FRB20220509G & 0.0894 & Y & & & & & & & & & & & \\
% FRB20190110C & 0.1224 & N & & & & & & & & & & & \\
% FRB20210603A & 0.177 & Y & & & & & & & & & & & \\
\bottomrule
\end{tabular}
\end{table*}
\FloatBarrier

\subsubsection{FRB pulse properties}\label{sec:frbprop}

We consider three sets of FRB properties: the scattering time $\tau_{\rm obs}$, the absolute rotation measure |$RM_{\rm ex}$|, and the linear and circular polarisation fractions. \citet{Scott2025} presents the methodology for derived $\tau_{\rm obs}$ values for CRAFT ICS FRBs. In brief, the approach taken by this work is to divide the bandwidth over which each FRB has significant power into four sub-channels, and fit each sub-channel's time series with a set of Gaussian burst profiles (each defined by a width, central time, and amplitude) alongside an exponential scattering term $\tau$. Scattering times are then scaled to a standard frequency of 1~GHz to facilitate comparisons between FRBs detected at different frequencies. We omit scattering times for FRBs (i.e., treat their values as unreliable) where large/unrealistic parameter values are found, namely where $\alpha$ {\myedit (the frequency dependence; where $\tau \sim \nu^{\alpha}$)} is consistent with a value of zero within the quoted 1$\sigma$ errors in \citet{Scott2025}. We are left with 28 FRBs from the CRAFT ICS sample (before further cuts due to measured galaxy properties; Section~\ref{sec:galprob}). For this work, we shift $\tau$ to the rest frame where we assume $\tau \propto \nu^{-4}$ and hence scale the observed scattering time by a factor of (1 + $z$)$^{3}$. As a sanity check, we test for correlations between scattering timescales in the rest frame with redshift, and find no correlation. %(Fig{\myedit ure}~\ref{fig:zdist}). 
We also find no correlation between scattering time and the dispersion measure contribution from the Milky Way for each FRB. We found estimates of the scattering time from the Milky Way, calculated from the YMW16 model \citep{Yao2017}, were negligible compared with all observed FRB scattering timescales, and so do not include them. 

Also presented by \citet{Scott2025} are the linear and circular polarisation fractions of the FRB pulses, and the extragalactic rotation measures RM$_{\rm ex}$ = RM$_{\rm obs}$ - RM$_{\rm MW}$, where a Milky Way (Galactic) contribution to the RM is subtracted from the observed $RM_{\rm obs}$. We then convert RM$_{\rm ex}$ to the rest frame (observed RM multiplied by a factor of (1+$z$)$^{2}$). The Galactic RM contribution model is derived from the \citet{Hutschenreuter2022} map via the {\sc frb} software package \citep{Prochaska2025}. These measurements are obtained from 1~MHz Stokes spectra integrated from Stokes I, Q, and U polarisation-calibrated dynamic spectra over each burst. A total of 32 FRBs in the CRAFT ICS sample have measured polarisation fractions, while 33 have RMs. Five of the CRAFT FRBs were detected outside the half power point in edge or corner beams on the ASKAP phased array feed, which can lead to significant residual polarisation calibration errors  \citep{Scott2025}. While we provide their observed polarisation fractions, we mark these values with an * in Table~\ref{tab:burstprop}, and do not include them in our analysis; results do not significantly change when they are included.
An additional four FRBs from the literature which we consider here have their RM reported. %, but these can only be compared with the stellar mass, SFR, and sSFR properties. 
%Across all these measured FRB properties with at least stellar mass or a {\sc galfit}-derived measurement available (see below), we have 40 FRBs. %(4 additional FRBs have only FRB burst properties available). 
Table~\ref{tab:burstprop} presents the scattering times, RMs, and polarisation fractions of the 44 FRBs considered in this work. %note 4 other FRBs do not have any host galaxy properties e.g. galactic plane or somesuch - just remove their rows from tables?

\newpage 

\subsubsection{Global host galaxy properties}\label{sec:galprob}

%\begin{figure}
%\centering
%\includegraphics[width=1.0\textwidth]{figures/z_log_Tau_obs_1GHz_1+z__3.png}%
%\caption{Scatter plot and correlation tests for scattering timescale in the rest frame of our sample of FRBs with their redshift. No correlation is seen.}
%\label{fig:zdist}
%\end{figure}

%It is important to use derived measurements of the FRB host galaxy that were made in a consistent and reliable manner. This is a motivation for this investigation into CRAFT ICS FRBs, where t
The majority of the FRB hosts considered here have global galaxy properties presented by \citet{Gordon2023} derived via spectral energy distribution modelling with {\sc Prospector} \citep{Johnson2021}. In that work, FRBs were only included if they had a high PATH \citep[Probabilitic Association of Transients to their Hosts;][]{Aggarwal2021} posterior probability of 90\%, were not near a bright foreground star, were detected in at least three optical/IR photometric bands alongside a good-quality optical spectrum (for reliable modelling), and had burst spectral energies above 10$^{27}$~erg\,Hz$^{-1}$ (to exclude low-energy bursts that would be missed at higher redshifts spanned by the sample). Some twenty-three FRB hosts were analysed in \citet{Gordon2023}, with 15 of these also having reliable scattering measurements as described above. The $R$-band AB magnitude, total stellar mass formed over the life of the galaxy $M_{\rm F}$, the present-day or `surviving' stellar mass $M_{*}$, the gas-phase metallicity log($Z_{\rm gas}$), the stellar metallicity log($Z_{\rm star}$), the integrated 0--100~Myr star formation rate (SFR), the integrated 0--100~Myr specific SFR (SFR/$M_{*}$, or sSFR), dust extinction due to old and young stars ($A_{\rm V,old}$, $A_{\rm V,young}$) and the mass-weighted age $t_{\rm m}$ from \citet{Gordon2023} are given in Table~\ref{tab:galprop}.  

One CRAFT ICS FRB (FRB\,20220610A), whose host galaxy was first reported by \citet{Ryder2023} and included a total stellar mass and stellar metallicity measure, had a few updated global galaxy properties (stellar mass, SFR, and mass-weighted age) presented by \citet{Gordon2024}, also derived through {\sc Prospector}, which are used in this analysis. Another CRAFT FRB (FRB\,20240201A) host galaxy has stellar mass and SFR values presented in \citet{Chang2015}. An additional eight FRBs and their host galaxy SFRs and stellar masses taken from the literature, which were also considered in the study by \citet{Acharya2025}, are included (see last 8 rows of Table~\ref{tab:galprop}). We additionally calculate their corresponding sSFR values.

Furthermore, H$\alpha$ fluxes \citep[a tracer of current star formation, specially $>10 M_{\odot}$ stars formed over the last $<20$~Myr;][]{Kennicutt1998} presented in \citet{Eftekhari2023} are considered. The same spectra also presented by \citet{Gordon2023} are used to derive %continuum fluxes at the location of the H$\alpha$ emission, and in turn 
H$\alpha$ equivalent widths (EWs), a measure of the current to past average star formation - a proxy for sSFR and recent star formation. In addition, [S\,{\sc ii}] line doublet ratios ([S\,{\sc ii}]$\lambda6716$/[S\,{\sc ii}]$\lambda6731$) are provided through the {F}ast and {U}nbiased F{RB} host galax{Y} (FURBY) program \citep{Muller2025}. The [S\,{\sc ii}] close doublet intensity ratio has been shown to be a robust diagnostic for electron density in ionised gaseous nebulae \citep{Wang2004}. A few other FRB host measurements including magnitude, SFR, and $H\alpha$ flux, as well as the SFR, stellar mass, and mass-weighted age for FRB\,20230708A in particular, are also provided by \citet{Muller2025}. %H-alpha is an observational proxy for mass-weighted mean age, in that greater star formation now relative to the recent past (high H-alpha EW) would correspond to a shorter mean age

Also considered for each FRB host is the optical disc semi-minor/semi-major axis ratio b/a, as well as the the inclination of the host's optical disc which is inferred -- with many caveats -- from b/a. \cite{Gordon2025} demonstrated that the majority of ASKAP FRBs are consistent with originating in a disc. Hosts with low values of b/a will on average have travelled through more disc material. We also considered the cosine of the inclination, but found similar correlation results to just using the inclination angle. This and the host galaxy's effective radius $R_{\rm eff}$ were measured through {\sc galfit} (Marnoch et al., in prep), as was the projected offset of the FRB localisation from the galaxy centre. This data was available for 28 (26 with inclination measurements) of our sample. Combining stellar masses with effective radii measures, we also consider the stellar gravitational potential ($M_{*}$/$R_{\rm eff}$) and the stellar surface density, or compactness ($M_{*}$/($R_{\rm eff}$)$^{2}$) of the host galaxy.  

Besides considering the whole FRB sample described above, we additionally investigate correlations for a subset of FRB host galaxies found to be actively star-forming (rather than transitioning or quiescent in the literature; we remove four of these FRB host galaxies in that analysis to check whether these rarer hosts conceal or drive any correlations). These values are presented in Table~\ref{tab:galfitprop}.

We note that these are \textit{global} host galaxy properties. % - any excess scattering time of an FRB could also be attributed to the local environment of the progenitor. 
FRB scattering is due to the integrated properties of the medium along a specific line of sight, including the local environment of the progenitor, and variations in the host ISM from its global mean. Detailed analysis of, for example, the local surface stellar density within the galaxy, or offsets from spiral arms rather than more simply the galaxy centre, is underway in separate studies \citep[][Mannings et al., in prep.]{Gordon2025}. 

\subsection{Correlations}\label{sec:corr}

To assess correlations between the scattering timescale measured and these global galaxy properties, we employ Spearman and Pearson coefficients. These values span a range of --1 to 1, where Spearman is a measure of the monotonicity of the relationship of the two parameters considered, while the Pearson coefficient examines the linearity. {\myedit In cases where we find statistically significant Pearson correlations but not for Spearman (e.g. Section~\ref{sec:scatter} for scattering timescale with potential), it highlights that the linear correlation between the two properties is more significant than any monotonic or rank correlation present in the data (which could be in some cases also influenced by relatively small sample sizes). By testing both measures we hence probe two different measures of potential correlations in our datasets.} We use the Python {\sc scipy} modules for these two correlation measures \citep{2020SciPy-NMeth}. For determining whether any correlation is significant, we derive p-values (henceforth $p$), where we consider $p \leq 0.01$ to reflect a highly significant correlation, and $0.01 < p \leq 0.05$ a weakly significant correlation. While these measures do not consider errors, we include the errors in previously described tables, and in figures showcasing any correlations (or lack thereof). %We use the {\sc scipy} module for these calculations }.

For Spearman coefficients, we use $p$-values supplied by the {\sc scipy} package. For the Pearson correlation coefficient, we instead shuffle the two sets of values being correlated 10,000 times and determine $p$ as the fraction of times the shuffled arrays return a stronger correlation than the non-shuffled dataset. We see similar results regardless of the choice of calculation method of $p$ for both correlation statistics. %, and highlight any meaningful differences in a correlation being deemed significant in Section~\ref{sec:results}. No meaningful difference actually found!
In addition, we perform bootstrapping 10,000 times per correlation through {\sc scipy} to determine the likelihood of obtaining a result with a strongly or weakly significant correlation (i.e. $p \leq 0.01$ or 0.05), to further assess how robust any one result may be. 

Lastly, to factor in measurement error of global galaxy properties, we do the following for any statistically significant result found. We resample each global galaxy property datapoint by drawing from a Gaussian distribution where the width is determined by the datapoint's error 1,000 times, calculate Spearman and Pearson correlations for each new distribution, and examine the resulting distribution of $p$-values. A higher fraction of $p < 0.01$ and 0.05 than otherwise indicates that such results are robust despite measurement error. 

\section{Results and Discussion}\label{sec:results}

Correlations, $p$, and bootstrap-derived percentages of correlations of host galaxy properties found to be strongly or weakly significant from 10,000 trials  probabilities are reported in Table~\ref{tab:correlations} for scattering timescales, rotation measure, and circular and linear polarisation fractions. We indicate the sample size for each correlation considered. For a few correlation results we also give the fraction of correlations with a $p$-value $< 0.01$ and 0.05 in both Spearman and Pearson after 1,000 resamples of global host galaxy property datasets when incorporating errors in Table~\ref{tab:errorcorrfraction}.

The following correlations are found which are, based on their corresponding $p$-value, considered to be statistically significant:

\begin{itemize}
    \item Highly significant correlations between scattering timescale and the mass-weighted age and stellar surface density (compactness).
    \item Weakly significant correlations between scattering timescale and gas-phase metallicity, $H\alpha$ EW, and gravitational potential.
    \item Highly significant anti-correlation between |$RM_{\rm ex}$| and the optical disc axis ratio b/a.
    %\item Weakly significant correlation between |$RM_{\rm ex}$| and compactness.
    \item Weakly significant anti-correlations between circular polarisation fraction with stellar mass, potential, compactness, and effective radius which are affected by one datapoint. 
\end{itemize}

We discuss possible explanations for significant correlations (or lack thereof) in Sections~\ref{sec:scatter}, \ref{sec:RM}, and \ref{sec:pol}, and discuss the significance and likelihood of these correlations in Section~\ref{sec:caveats}.

\begin{figure*}
\centering
\resizebox{.95\textwidth}{!}{%
\includegraphics[height=9cm]{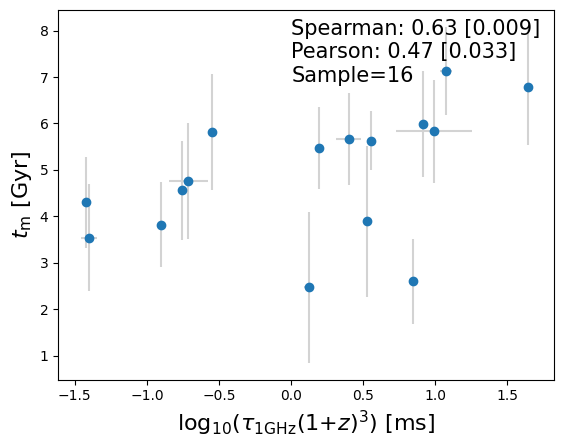}%
\includegraphics[height=9cm]{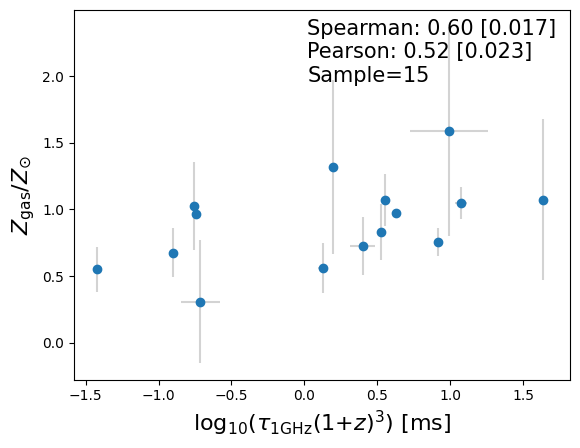}
}
\resizebox{.95\textwidth}{!}{%
\includegraphics[height=9cm]{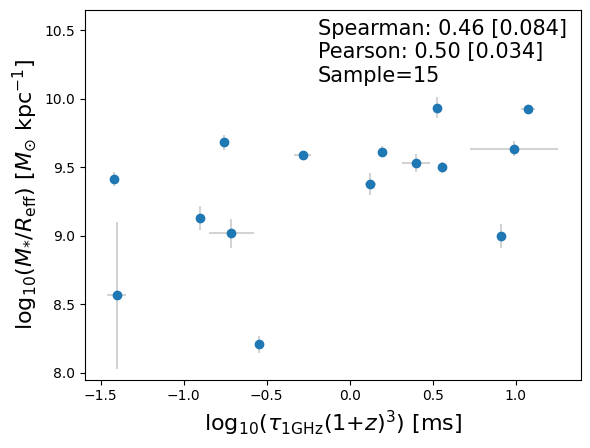}%
\includegraphics[height=9cm]{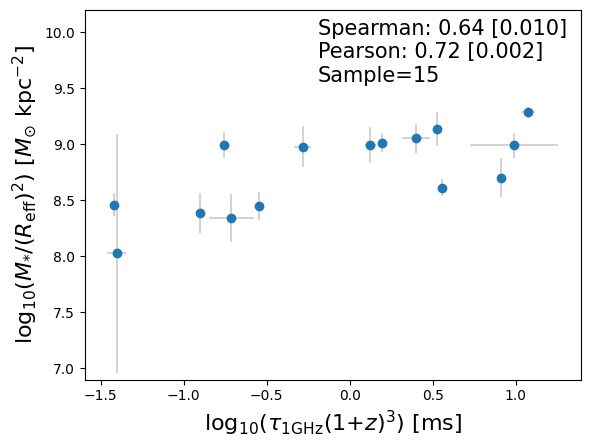}
}
\resizebox{.95\textwidth}{!}{%
\includegraphics[height=9cm]{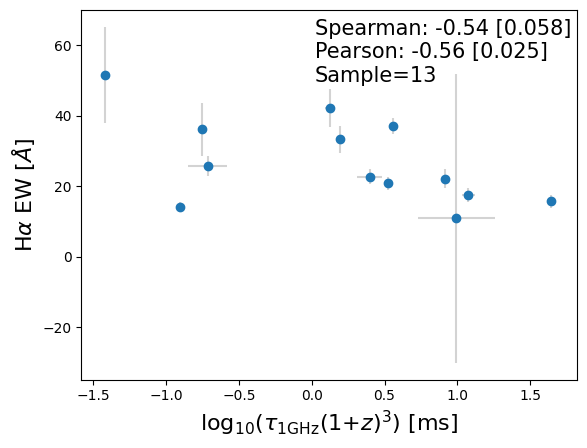}%
\includegraphics[height=9cm]{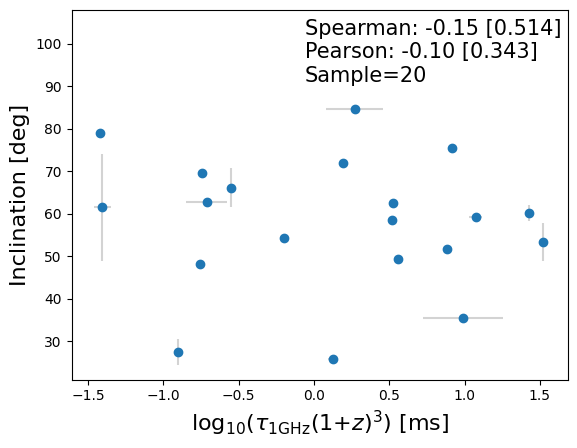}
}
\caption{Scatter plots for the logarithm of the rest-frame scattering time at a 1~GHz reference frequency with global galaxy properties: mass-weighted age, gas-phase metallicity, potential $M_{*}$/$R_{\rm eff}$, stellar surface density or compactness $M_{*}$/($R_{\rm eff}$)$^{2}$, $H\alpha$ EW, and optical galaxy inclination angle. Spearman and Pearson correlation coefficients, accompanied by p-values in square brackets and sample size, are in the upper-right legend.}
\label{fig:taucorr}
\end{figure*}

\begin{figure*}
\ContinuedFloat
\centering
\resizebox{.95\textwidth}{!}{%
\includegraphics[height=9cm]{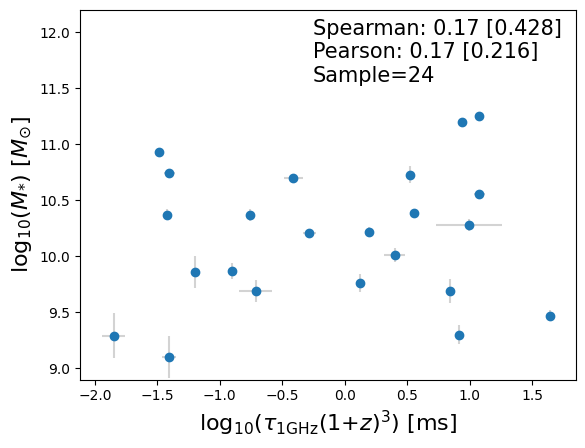}%
\includegraphics[height=9cm]{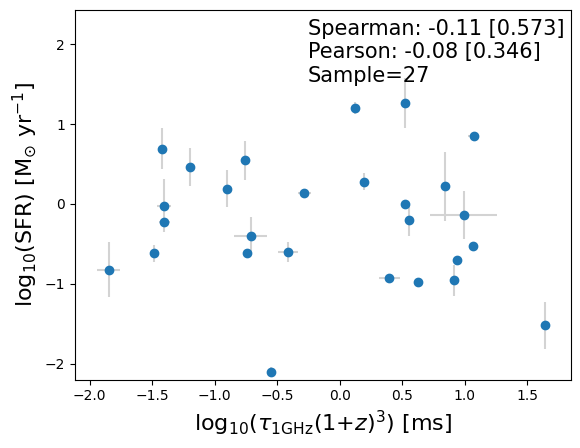}
}
\caption{Continued. Two scatter plots for the logarithm of the rest-frame scattering time at a 1~GHz reference frequency with global galaxy properties: current-day stellar mass and SFR.}
\end{figure*}

%\begin{figure*}
%\centering
%\resizebox{.95\textwidth}{!}{%
%\includegraphics[height=9cm]{figures/tau/log_Tau_obs_1GHz_1+z__3_log_SFR.png}%
%\includegraphics[height=9cm]%{figures/tau/log_Tau_obs_1GHz_1+z__3_log_SFR_M.png}
%}
%\caption{}
%\label{fig:taucorr2}
%\end{figure*}

\subsection{Scattering timescale correlations}\label{sec:scatter}

We present a few scatter plots for the range of global host galaxy parameters with the rest-frame $\tau_{\rm obs}$ in Figure~\ref{fig:taucorr}, when considering the whole sample with reliable scattering measurements. For the majority of the global galaxy properties, we find no correlation with the scattering timescale, including for stellar mass measurements or star formation (Table~\ref{tab:correlations}). This result, consistent with the findings by \citet{Acharya2025}, % which included a few non-starforming FRB host galaxies, 
implies that the size or mass of the galaxy, or how much star formation is happening throughout the whole galaxy, does not significantly impact the FRB pulse travelling through the FRB sight line. We note that scattering of the FRB signal limits the FRBs we can detect (as strongly scattered FRBs may smear out the signal and hence reduce the S/N below detection thresholds). Therefore, we may be missing any FRBs that are strongly scattered by high stellar masses, SFR, etc., and are only probing the closer `surface' of each FRB host galaxy. That said, we have no evidence for missing FRBs due to this effect. If we were missing these FRBs, such hosts with FRBs we cannot detect could have masses or SFRs above some critical value, which may correspond to more extreme cases not covered by current FRB host populations; see  \citet{Gordon2023}.

We highlight a particular absence of any correlation with galaxy inclination angle, as well as the b/a ratio (Marnoch et al., in prep). This is in tension with the result reported by \citet{Bhardwaj2024}, of a substantial selection bias against detecting FRBs in galaxies with large inclination angles. If this were true, then combined with a simplistic expectation of greater scattering times for FRBs travelling through more of its host galaxy, we would expect on average a larger scattering time for FRBs within edge-on galaxies (unless they all were on the `near' side of edge-on galaxies), and correspondingly less scattering from FRBs within face-on spirals \citep[see also Fig{\myedit ure}~2 and discussion by][]{Kovacs2024}. Given the discrepancy, this finding would benefit from greater statistics and scrutiny into galaxy inclination measurements. We leave further analysis on the effect of galaxy inclination on FRB detection to a separate study (Marnoch et al., in prep.).

While a Spearman and Pearson correlation coefficient of --0.50 and --0.51 is found for scattering timescale with the [S\,{\sc ii}] close doublet line ratio, this cannot be deemed significant due to the small sample of 5 FRBs with reliable scattering measurements ($p = 0.391$ and 0.142 respectively). A much larger sample is required to see if such line doublet ratios can reveal a trend with electron density that would affect scattering timescales. 

\subsubsection{Scattering timescale dependence on compactness and gravitational potential}\label{sec:compactness}

We find a Spearman correlation coefficient (0.64) and Pearson correlation coefficient (0.72) with the stellar surface density, or compactness, $M_{*}$/($R_{\rm eff}$)$^{2}$. Corresponding $p$-values are 0.010 and 0.001 respectively. A weaker significance Pearson correlation is also found for a proxy of the galaxy potential, $M_{*}$/$R_{\rm eff}$, of 0.50 ($p = 0.034$). For compactness, in 10,000 iterations of bootstrapping our sample, we find a Spearman correlation with $p < 0.01$ is found 52.4\% of the time, and a Spearman correlation with $p < 0.05$ is seen for 76.8\% of our bootstrapped tests. For Pearson, these are 72.9\% ($p < 0.01$) and 90.2\% ($p < 0.05$). Lastly, when considering resampling based on error measurements (Table~\ref{tab:errorcorrfraction}), $p < 0.01$ [0.05] is found for Spearman and Pearson correlations 9.4\% [92.6\%] and 87.9\% [100\%] of the time respectively. This suggests that this is a reasonably robust result, rather than one arising by chance due to measurement error, or from any outliers.

Taken at face value, this correlation implies that the ionised gas density scales with the stellar surface density, a result that is not wholly surprising. {\myedit While this is not directly found in previous studies, \citet{Shimakawa2015} and \citet{Reddy2023} did find correlations between SFR surface density with electron density; these two surface density measures are related through the resolved star forming
main sequence \citep[e.g.][]{Baker2022}.} The weaker correlation with galaxy gravitational potential, if real, also aligns with this result. %The lack of correlation found with just the stellar mass highlights that the distribution of material in more massive galaxies needs to be considered (which does not account for the spread of this material in the FRB host). 
However, we highlight the sample size of 15 here is modest (see also Section~\ref{sec:caveats}).

\subsubsection{Scattering timescale dependence on mass-weighted age and gas-phase metallicity}

\begin{table*}[h!]
\centering
\caption{Results of our correlation tests for the rest-frame scattering time and absolute value of the rotation measure. For each galaxy property we give the Spearman correlation coefficient and corresponding $p$, the Pearson correlation value and corresponding $p$, the sample size $N$, the percentage of bootstrapped Spearman correlation coefficients found with $p < 0.01$ and $0.05$ (strong and weak correlation respectively), and likewise for Pearson. {\myedit To aid the reader's eye, we highlight in {\bf bold} cases where a low $p$ value is seen but a bootstrap analysis indicates that a weak correlation arises at least 50\% of the time.}}
\addtolength{\tabcolsep}{-0.2em}
\begin{tabular}{>{\rowmac}l>{\rowmac}l|>{\rowmac}r>{\rowmac}r>{\rowmac}r>{\rowmac}r>{\rowmac}r>{\rowmac}c>{\rowmac}c>{\rowmac}c>{\rowmac}c<{\clearrow}}
\toprule
 FRB measure & Galaxy measure & Spearman & p-value & Pearson & p-value & $N$ & S($p < 0.01$) \% & S($p < 0.05$) \% & P($p < 0.01$) \% & P($p < 0.05$)\% \\
\midrule
  log($\tau_{\rm 1 GHz}$*(1+$z$)$^{3}$) & AB Mag &  --0.37 &  0.191 & --0.49 &  0.035 &  14 &  12.7 &  28.5 & 18.6 & 42.1 \\
 & $A_{\rm V,o}$ & --0.01 & 0.982 & 0.12 & 0.353 & 14 & 1.2 & 4.4 & 0.3 & 1.5 \\
 & $A_{\rm V,y}$ & 0.13 & 0.656 & 0.05 & 0.443 & 14 & 1.2 & 6.0 & 1.6 & 6.8 \\
 & $\frac{Z_{\rm star}}{Z_{\odot}}$ & 0.05 & 0.869 & 0.13 & 0.316 & 15 & 2.1 & 7.3 & 1.7 & 6.0 \\
\setrow{\bfseries}  &  \textbf{$\frac{Z_{\rm gas}}{Z_{\odot}}$} & 0.60 & 0.017 & 0.52 & 0.023 & 15 & 40.0 & 68.0 & 19.8 & 54.9 \\
 & log(SFR) & --0.11 & 0.573 & --0.08 & 0.346 & 27 & 2.8 & 9.7 & 1.8 & 6.7 \\
 & log($M_{\rm F}$) & 0.11 & 0.694 & 0.13 & 0.325 & 15 & 4.0 & 11.1 & 4.6 & 11.9 \\
 & log($M_{*}$) & 0.17 & 0.428 & 0.17 & 0.216 & 24 & 6.3 & 16.6 & 4.0 & 13.3 \\
 & log($\frac{SFR}{M}$) & --0.27 & 0.203 & --0.23 & 0.283 & 24 & 10.6 & 25.0 & 6.9 & 19.6 \\
\setrow{\bfseries} & \textbf{$t_{\rm m}$} & 0.63 & 0.009 & 0.47 & 0.033 & 16 & 53.1 & 73.7 & 24.2 & 46.2 \\
 & $R_{\rm eff}$ & --0.21 & 0.342 & --0.26 & 0.122 & 22 & 8.1 & 20.1 & 9.4 & 23.5 \\
 & log($R_{\rm eff}$) & --0.21 & 0.342 & --0.16 & 0.242 & 22 & 7.8 & 19.8 & 7.8 & 16.7 \\
 & Offset & --0.01 & 0.950 & --0.16 & 0.243 & 22 & 1.5 & 6.1 & 2.7 & 10.4 \\
 & $\frac{Offset}{R_{\rm eff}}$ & 0.15 & 0.497 & 0.10 & 0.337 & 22 & 5.1 & 13.5 & 3.7 & 11.5 \\
 & Inclination & --0.15 & 0.514 & --0.10 & 0.343 & 20 & 2.9 & 10.1 & 1.4 & 5.8 \\
 & b/a & 0.17 & 0.451 & 0.11 & 0.309 & 21 & 3.3 & 11.6 & 0.9 & 4.5 \\
 & $F_{\rm H}$\textsubscript{$\alpha$} & 0.25 & 0.324 & 0.38 & 0.058 & 18 & 7.5 & 19.7 & 10.8 & 29.9 \\
\setrow{\bfseries}  &  \textbf{$H\alpha$ EW}  & --0.54 & 0.058 & --0.56 & 0.026 & 13 & 32.2 & 50.0 & 31.6 & 52.9 \\
\setrow{\bfseries} & \textbf{log($\frac{M_{*}}{R_{\rm eff}}$)} & 0.46 & 0.084 & 0.50 & 0.034 & 15 & 20.2 & 41.5 & 25.1 & 51.6 \\
\setrow{\bfseries} & \textbf{log($\frac{M_{*}}{R_{\rm eff}^{2}}$)} & 0.64 & 0.010 & 0.72 & 0.001 & 15 & 52.4 & 76.8 & 72.9 & 90.2 \\
 & [S\,{\sc ii}] ratio & --0.50 & 0.391 & --0.51 & 0.142 & 5 & 24.2 & 32.0 & 14.3 & 17.4 \\
  %&  $C_{\rm H}$\textsubscript{$\alpha$} & 0.15 &  0.616 &  0.31 &  0.136 &  13 & 4.6 &  11.9 &  2.0 &  8.1 \\
  \hline
  log(|$RM_{\rm ex}$|*(1+$z$)$^{3}$) & AB Mag & 0.07 & 0.726 & 0.04 & 0.412 & 26 & 2.1 & 7.9 & 0.8 & 3.9 \\
 & $A_{\rm V,o}$ & 0.05 & 0.852 & 0.25 & 0.168 & 17 & 2.1 & 7.6 & 6.4 & 16.7 \\
 & $A_{\rm V,y}$ & 0.08 & 0.757 & 0.18 & 0.254 & 17 & 1.8 & 6.8 & 0.7 & 4.4 \\
 & $\frac{Z_{\rm star}}{Z_{\odot}}$ & 0.10 & 0.695 & 0.05 & 0.416 & 18 & 1.7 & 6.6 & 3.2 & 9.2 \\
 & $\frac{Z_{\rm gas}}{Z_{\odot}}$ & --0.20 & 0.403 & --0.19 & 0.215 & 19 & 1.5 & 9.0 & 0.1 & 2.1 \\
 & log(SFR) & --0.05 & 0.811 & 0.01 & 0.487 & 29 & 1.6 & 5.6 & 0.3 & 2.0 \\
 & log($M_{\rm F}$) & 0.04 & 0.884 & 0.03 & 0.454 & 18 & 0.5 & 3.2 & 0.3 & 1.6 \\
 & log($M_{*}$) & 0.11 & 0.607 & 0.12 & 0.280 & 24 & 0.5 & 3.8 & 0.2 & 2.0 \\
 & log($\frac{SFR}{M}$) & 0.05 & 0.834 & 0.02 & 0.467 & 24 & 0.8 & 3.7 & 0.1 & 0.9 \\
 & $t_{\rm m}$ & --0.05 & 0.836 & --0.13 & 0.303 & 19 & 1.4 & 5.7 & 1.0 & 4.7 \\
 & $R_{\rm eff}$ & --0.12 & 0.566 & --0.20 & 0.173 & 26 & 2.0 & 8.4 & 1.2 & 8.0 \\
 & log($R_{\rm eff}$) & --0.12 & 0.566 & --0.11 & 0.298 & 26 & 1.8 & 8.0 & 0.9 & 5.6 \\
 & Offset & 0.04 & 0.844 & 0.03 & 0.443 & 26 & 2.2 & 7.2 & 2.2 & 7.4 \\
 & $\frac{Offset}{R_{\rm eff}}$ & 0.17 & 0.397 & 0.05 & 0.420 & 26 & 6.7 & 17.1 & 8.0 & 16.0 \\
 & Inclination & 0.25 & 0.233 & 0.37 & 0.040 & 24 & 5.8 & 19.1 & 15.8 & 40.2 \\
\setrow{\bfseries}  & \textbf{b/a} & --0.41 & 0.038 & --0.53 & 0.002 & 26 & 30.0 & 53.4 & 61.9 & 83.7 \\
 & $F_{\rm H}$\textsubscript{$\alpha$} & --0.08 & 0.736 & --0.10 & 0.350 & 22 & 1.2 & 5.3 & 0.5 & 2.2 \\
 & $H\alpha$ EW & 0.26 & 0.333 & 0.09 & 0.366 & 16 & 10.0 & 20.5 & 3.6 & 10.1 \\
 & log($\frac{M_{*}}{R_{\rm eff}}$) & 0.13 & 0.606 & 0.20 & 0.223 & 17 & 1.1 & 4.9 & 0.9 & 4.9 \\
 & log($\frac{M_{*}}{R_{\rm eff}^{2}}$) & 0.36 & 0.153 & 0.34 & 0.091 & 17 & 9.4 & 27.7 & 3.8 & 18.4 \\
 & [S\,{\sc ii}] ratio & 0.37 & 0.468 & 0.44 & 0.185 & 6 & 23.6 & 30.0 & 24.1 & 32.7 \\
  \bottomrule
\end{tabular}
\label{tab:correlations}
\end{table*}

\begin{table*}[h!]\ContinuedFloat
\centering
\caption{Continued, for linear and circular polarisation fractions.}
\addtolength{\tabcolsep}{-0.2em}
\begin{tabular}{>{\rowmac}l>{\rowmac}l|>{\rowmac}r>{\rowmac}r>{\rowmac}r>{\rowmac}r>{\rowmac}r>{\rowmac}c>{\rowmac}c>{\rowmac}c>{\rowmac}c<{\clearrow}}
\toprule
 FRB measure & Galaxy measure & Spearman & p-value & Pearson & p-value & N & S($p < 0.01$) \% & S($p < 0.05$) \% & P($p < 0.01$) \% & P($p < 0.05$)\% \\
\midrule
Linear  & AB Mag &0.35 & 0.247 & 0.46 & 0.066 & 13 & 16.7 & 29.6 & 20.4 & 38.8 \\
polarisation   & $A_{\rm V,o}$ & --0.32 & 0.292 & --0.42 & 0.088 & 13 & 12.0 & 23.0 & 17.5 & 32.9 \\
fraction  & $A_{\rm V,y}$ & --0.25 & 0.411 & --0.33 & 0.132 & 13 & 6.5 & 17.2 & 7.1 & 20.4 \\
 & $\frac{Z_{\rm star}}{Z_{\odot}}$ & 0.17 & 0.560 & 0.26 & 0.177 & 14 & 4.6 & 12.1 & 4.0 & 12.8 \\
 & $\frac{Z_{\rm gas}}{Z_{\odot}}$ & --0.16 & 0.562 & --0.37 & 0.101 & 15 & 4.4 & 12.6 & 9.5 & 26.3 \\
 & log(SFR) & -0.07 & 0.770 & --0.22 & 0.362 & 19 & 4.4 & 11.8 & 6.2 & 16.8 \\
 & log($M_{\rm F}$) & 0.05 & 0.863 & 0.02 & 0.472 & 14 & 3.1 & 9.2 & 1.6 & 5.2 \\
 & log($M_{*}$) & --0.12 & 0.673 & --0.17 & 0.274 & 15 & 3.4 & 10.8 & 0.8 & 5.1 \\
 & log($\frac{SFR}{M}$) & 0.06 & 0.845 & --0.18 & 0.263 & 15 & 4.7 & 12.0 & 14.2 & 26.4 \\
 & $t_{\rm m}$ & 0.02 & 0.934 & 0.05 & 0.424 & 15 & 4.8 & 12.2 & 2.1 & 7.3 \\
 & $R_{\rm eff}$ & 0.01 & 0.964 & --0.04 & 0.406 & 22 & 3.1 & 10.0 & 3.6 & 12.3 \\
 & log($R_{\rm eff}$) & 0.01 & 0.964 & 0.01 & 0.470 & 22 & 3.0 & 9.3 & 3.1 & 9.5 \\
 & Offset & --0.01 & 0.966 & 0.03 & 0.459 & 22 & 2.6 & 8.5 & 1.0 & 4.1 \\
 & $\frac{Offset}{R_{\rm eff}}$ & --0.09 & 0.687 & --0.11 & 0.296 & 22 & 3.9 & 10.7 & 3.0 & 8.5 \\
 & Inclination & 0.15 & 0.510 & 0.26 & 0.130 & 21 & 3.4 & 10.5 & 9.5 & 22.2 \\
 & b/a & --0.04 & 0.859 & 0.02 & 0.464 & 22 & 2.2 & 7.3 & 4.3 & 12.9 \\
 & $F_{\rm H}$\textsubscript{$\alpha$}& --0.24 & 0.457 & --0.50 & 0.069 & 12 & 9.1 & 18.7 & 16.0 & 38.2 \\
 & $H\alpha$ EW & --0.11 & 0.729 & 0.13 & 0.354 & 12 & 3.5 & 12.2 & 1.6 & 5.1 \\
 & log($\frac{M_{*}}{R_{\rm eff}}$) & --0.18 & 0.543 & --0.14 & 0.308 & 14 & 6.8 & 15.6 & 4.6 & 12.4 \\
 & log($\frac{M_{*}}{R_{\rm eff}^{2}}$) & --0.16 & 0.584 & --0.06 & 0.419 & 14 & 11.2 & 15.3 & 12.4 & 21.7 \\
 & [S\,{\sc ii}] ratio & 0.20 & 0.800 & --0.25 & 0.421 & 4 & 46.6 & 46.6 & 42.8 & 47.3 \\
  \hline
 Circular  & AB Mag & 0.35 & 0.244 & 0.28 & 0.166 & 13 & 10.5 & 24.6 & 5.9 & 17.6 \\
polarisation & $A_{\rm V,o}$ & 0.03 & 0.929 & --0.07 & 0.481 & 13 & 2.3 & 6.5 & 3.4 & 7.2 \\
fraction & $A_{\rm V,y}$ & 0.30 & 0.327 & 0.42 & 0.083 & 13 & 7.7 & 19.5 & 9.2 & 28.2 \\
& $\frac{Z_{\rm star}}{Z_{\odot}}$ & --0.12 & 0.679 & --0.21 & 0.253 & 14 & 2.4 & 8.2 & 0.8 & 5.2 \\
 & $\frac{Z_{\rm gas}}{Z_{\odot}}$ & --0.08 & 0.763 & 0.15 & 0.223 & 15 & 4.5 & 11.5 & 3.4 & 6.6 \\
 & log(SFR) & 0.15 & 0.545 & --0.24 & 0.313 & 19 & 6.3 & 14.9 & 5.9 & 19.6 \\
 & log($M_{\rm F}$) & --0.29 & 0.312 & --0.24 & 0.202 & 14 & 10.1 & 21.2 & 3.1 & 11.8 \\
\setrow{\bfseries} & \textbf{log($M_{*}$)} & --0.39 & 0.145 & --0.58 & 0.022 & 15 & 17.3 & 34.2 & 38.8 & 57.6 \\
 & log($\frac{SFR}{M}$) & 0.30 & 0.276 & 0.13 & 0.292 & 15 & 2.2 & 12.6 & 0.6 & 2.3 \\
 & $t_{\rm m}$ & --0.33 & 0.235 & 0.01 & 0.503 & 15 & 13.6 & 27.6 & 2.6 & 8.5 \\
 & $R_{\rm eff}$ & --0.07 & 0.742 & --0.24 & 0.140 & 22 & 1.8 & 7.4 & 3.6 & 15.5 \\
\setrow{\bfseries} & \textbf{log($R_{\rm eff}$)} & --0.07 & 0.742 & --0.47 & 0.024 & 22 & 2.1 & 7.9 & 36.0 & 56.0 \\
 & Offset & --0.12 & 0.595 & 0.03 & 0.400 & 22 & 5.9 & 14.3 & 16.5 & 30.6 \\
 & $\frac{Offset}{R_{\rm eff}}$ & --0.05 & 0.811 & 0.24 & 0.132 & 22 & 4.2 & 11.5 & 17.9 & 29.4 \\
 & Inclination & --0.24 & 0.300 & --0.17 & 0.237 & 21 & 4.8 & 15.8 & 5.2 & 16.8 \\
 & b/a & 0.27 & 0.221 & 0.20 & 0.186 & 22 & 6.5 & 20.4 & 6.1 & 19.8 \\
 & $F_{\rm H}$\textsubscript{$\alpha$}& --0.19 & 0.554 & 0.07 & 0.269 & 12 & 6.2 & 14.6 & 4.7 & 11.1 \\
 & $H\alpha$ EW & 0.09 & 0.777 & 0.00 & 0.448 & 12 & 6.9 & 12.2 & 1.6 & 3.9 \\
 & log($\frac{M_{*}}{R_{\rm eff}}$) & --0.44 & 0.112 & --0.48 & 0.046 & 14 & 19.0 & 36.7 & 22.4 & 42.7 \\
 & log($\frac{M_{*}}{R_{\rm eff}^{2}}$)) & --0.51 & 0.062 & --0.20 & 0.249 & 14 & 25.8 & 46.9 & 4.1 & 9.3 \\
 & [S\,{\sc ii}] ratio & --0.32 & 0.684 & --0.79 & 0.171 & 4 & 27.6 & 27.6 & 32.4 & 40.9 \\
  \bottomrule
\end{tabular}
\end{table*}

\begin{table*}[h!]\ContinuedFloat
\centering
\caption{Continued, for total polarisation fractions.}
\addtolength{\tabcolsep}{-0.2em}
\begin{tabular}{>{\rowmac}l>{\rowmac}l|>{\rowmac}r>{\rowmac}r>{\rowmac}r>{\rowmac}r>{\rowmac}r>{\rowmac}c>{\rowmac}c>{\rowmac}c>{\rowmac}c<{\clearrow}}
\toprule
 FRB measure & Galaxy measure & Spearman & p-value & Pearson & p-value & N & S($p < 0.01$) \% & S($p < 0.05$) \% & P($p < 0.01$) \% & P($p < 0.05$)\% \\
\midrule
 Total  & AB Mag & 0.34 & 0.263 & 0.45 & 0.073 & 13 & 15.8 & 28.4 & 19.6 & 37.4 \\
polarisation & $A_{\rm V,o}$ & --0.29 & 0.329 & --0.43 & 0.084 & 13 & 10.0 & 19.8 & 16.9 & 32.7 \\
fraction & $A_{\rm V,y}$ & --0.25 & 0.401 & --0.31 & 0.151 & 13 & 6.3 & 17.1 & 6.0 & 17.3 \\
 & $\frac{Z_{\rm star}}{Z_{\odot}}$ & 0.19 & 0.512 & 0.26 & 0.185 & 14 & 5.2 & 13.1 & 4.1 & 13.0 \\
 & $\frac{Z_{\rm gas}}{Z_{\odot}}$ & --0.15 & 0.589 & --0.35 & 0.125 & 15 & 3.6 & 11.6 & 7.5 & 22.7 \\
 & log(SFR) & --0.12 & 0.616 & --0.23 & 0.168 & 19 & 5.2 & 14.1 & 7.3 & 18.9 \\
 & log($M_{\rm F}$) & 0.07 & 0.823 & 0.02 & 0.470 & 14 & 3.0 & 8.6 & 1.5 & 4.8 \\
 & log($M_{*}$) & --0.18 & 0.532 & --0.21 & 0.229 & 15 & 5.0 & 13.8 & 1.3 & 7.8 \\
 & log($\frac{SFR}{M}$) & 0.07 & 0.810 & --0.15 & 0.300 & 15 & 4.4 & 11.3 & 14.0 & 25.7 \\
 & $t_{\rm m}$ & 0.07 & 0.800 & 0.06 & 0.411 & 15 & 3.9 & 11.0 & 1.9 & 6.7 \\
 & $R_{\rm eff}$ & --0.01 & 0.966 & --0.06 & 0.382 & 22 & 3.6 & 10.5 & 4.0 & 11.8 \\
 & log($R_{\rm eff}$) & --0.01 & 0.966 & --0.03 & 0.452 & 22 & 2.9 & 9.8 & 4.2 & 11.9 \\
 & Offset & --0.05 & 0.818 & 0.03 & 0.477 & 22 & 2.8 & 9.5 & 1.0 & 4.8 \\
 & $\frac{Offset}{R_{\rm eff}}$ & --0.12 & 0.587 & --0.10 & 0.316 & 22 & 4.7 & 12.7 & 3.6 & 8.5 \\
 & Inclination & 0.13 & 0.584 & 0.24 & 0.148 & 21 & 3.4 & 9.6 & 8.3 & 20.3 \\
 & b/a & --0.02 & 0.933 & 0.04 & 0.429 & 22 & 2.4 & 7.7 & 4.8 & 13.5 \\
 & $F_{\rm H}$\textsubscript{$\alpha$}& --0.22 & 0.484 & --0.48 & 0.091 & 12 & 7.9 & 17.4 & 14.1 & 34.2 \\
 & $H\alpha$ EW & --0.13 & 0.681 & 0.14 & 0.362 & 12 & 5.7 & 13.5 & 1.8 & 5.2 \\
 & log($\frac{M_{*}}{R_{\rm eff}}$) & --0.21 & 0.474 & --0.18 & 0.270 & 14 & 6.6 & 15.8 & 4.7 & 13.3 \\
 & log($\frac{M_{*}}{R_{\rm eff}^{2}}$) & --0.17 & 0.563 & --0.07 & 0.423 & 14 & 10.1 & 15.3 & 10.9 & 19.8 \\
 & [S\,{\sc ii}] ratio & --0.32 & 0.684 & --0.79 & 0.171 & 4 & 27.6 & 27.6 & 32.4 & 40.9 \\
  \bottomrule
\end{tabular}
\end{table*}

\begin{table*}[h!]
\centering
\caption{Fraction of correlations with a corresponding $p$-value below 0.01 or 0.05 when resampling global galaxy properties based on measurement error 1,000 times.}
\addtolength{\tabcolsep}{-0.2em}
\begin{tabular}{llllll}
\toprule
 %FRB measure & Galaxy measure & Fraction Spearman $p < 0.01$ & Fraction Spearman $p < 0.05$ & Fraction Pearson $p < 0.01$ & Fraction Pearson $p < 0.05$ \\
  FRB  & Galaxy  & Fraction Spearman & Fraction Spearman & Fraction Pearson & Fraction Pearson \\
  measure & measure &  $p < 0.01$ & $p < 0.05$ &  $p < 0.01$ & $p < 0.05$ \\
\midrule
 log($\tau_{\rm 1 GHz}$*(1+$z$)$^{3}$) & log($\frac{M_{*}}{R_{\rm eff}}$) & 0.001 & 0.142 & 0.033 & 0.770 \\
 log($\tau_{\rm 1 GHz}$*(1+$z$)$^{3}$) & log($\frac{M_{*}}{R_{\rm eff}^{2}}$) & 0.136 & 0.555 & 0.583 & 0.805 \\
 log($\tau_{\rm 1 GHz}$*(1+$z$)$^{3}$) & $t_{\rm m}$ &  0.097 & 0.394 & 0.077 & 0.342 \\
 log($\tau_{\rm 1 GHz}$*(1+$z$)$^{3}$) & $\frac{Z_{\rm gas}}{Z_{\odot}}$ & 0.100 & 0.301 & 0.112 & 0.364 \\
 log($\tau_{\rm 1 GHz}$*(1+$z$)$^{3}$) & $H\alpha$ EW &  0.005 & 0.217 & 0.018 & 0.497 \\
 log(|($RM_{\rm ex}$)|) & b/a & 0.097 & 0.899 & 0.894 & 1.000 \\
  \bottomrule
\end{tabular}
\label{tab:errorcorrfraction}
\end{table*}

A few other properties are found to have a correlation with the rest-frame scattering time. The first is with the stellar mass-weighted age $t_{\rm m}$, i.e. the average age of stars in a galaxy. We find a Spearman correlation coefficient of 0.63 with corresponding $p = 0.009$ found across the full dataset considered here. This strong correlation also holds when we use less robust scattering timescale measurements which we previously excluded (Section~\ref{sec:frbsample}). The Pearson coefficient indicates a weak correlation as well; 0.47 ($p = 0.033$). From the top left plot of Figure~\ref{fig:taucorr} we see three outlier points, two with slightly larger error bars than the rest of the small sample of 15 - that said, the errors for most points span $>0.8$~Gyr and are not insignificant (we remind the reader that neither of our correlation measures consider errors). Nonetheless, in 10,000 iterations of bootstrapping our sample, we find a Spearman correlation with $p < 0.01$ is found 53.1\% of the time, and a Spearman correlation with $p < 0.05$ is seen for 73.7\% of our bootstrapped tests. When performing correlations on 1,000 resampled distributions based on the measurement error, we find 39.4\% of the time that $p < 0.05$ (Table~\ref{tab:errorcorrfraction}). %This suggests that this is a reasonably robust result, rather than a result arising by chance due to measurement error, or from any outliers.

It is not immediately clear why scattering time would increase solely due to a larger stellar mass-weighted age - the implication is that the older the stars in the host galaxy of the FRB, the more scattering that occurs of the FRB signal. This is not apparently driven by more massive host galaxies given the lack of a statistically significant correlation with stellar mass measures. We find a $p = 0.023$ supporting a Spearman correlation coefficient (0.60) with $t_{\rm m}$ even for the star-forming subset of FRB host galaxies (sample size of 14). 

%Perhaps then it relates to another attribute of host galaxies involving older stars. \citet{Jimenez2007} investigated spectra for over 20,000 early-type galaxies and found when considering mass-weighted ages that ``the more massive galaxies are not only the ones that contain the oldest stars, but are also more metal-rich'' (section 7). Could galaxy metallicity be a factor?

Another correlation found for scattering is with the gas-phase metallicity (top-right panel of Fig{\myedit ure}~\ref{fig:taucorr}); we find a weaker but statistically significant Spearman correlation coefficient of 0.60 ($p = 0.017$). With bootstrapping, we find 40.0\% of the time that correlations have $p < 0.01$, and 68.0\% of cases that $p < 0.05$. Error bars are large (the $Z_{\rm gas}$/$Z_{\odot}$ error is $ > 1$ for five datapoints out of 16), and we note that ten of the points would be consistent within the 1$\sigma$ errors with a flat line (no correlation) at $Z_{\rm gas}$/$Z_{\odot}$ = 0.7. When resampling datapoints based on the errors, a Pearson correlation with $p < 0.05$ is found 34.2\% of the time. We highlight that in contrast no correlation is found with the stellar metallicity; whereas higher metallicity \textit{gas} is a potential driver for increased scattering time.

\citet{Jimenez2007} investigated spectra for over 20,000 early-type galaxies and found when considering mass-weighted ages that ``the more massive galaxies are not only the ones that contain the oldest stars, but are also more metal-rich'' (section 7). If we presume that this weakly significant positive correlation with gas-phase metallicity is real, then there may be a connection between the two properties for our sample. The older the stellar population, generally the higher the metallicity, so the finding between $\tau$ with both mass-weighted age and gas-phase metallicity may be tracing the same effect. Gas-phase metallicity has been found  by \cite[e.g.][]{Mingozzi2020, Grasha2022, Ji2022} to correlate with the ionisation parameter (the ratio of the number density of incident ionising photons and the number density of hydrogen atoms), or the `hardness' of the radiation field within the host galaxy. Thus, the higher the gas-phase metallicity of the FRB host galaxy, the more ionising photons and potentially {\myedit free} electrons within the galaxy which could increase burst scattering. Another effect is that metal-rich gas has been found to be cooler than metal-poor gas in galaxies. It is possible that such galaxies %have more cold molecular gas clouds --- ergo, a clumpier interstellar medium 
have a harder radiation field, and hence an increase in the ionised baryon fraction which could further contribute to the correspondingly larger scattering timescale.

\citet{Sharma2024} highlighted a deficit of low-mass FRB hosts compared with the occurrence of star formation in the Universe. This study argued that FRBs are a biased tracer of star formation, with the bias driven by galaxy metallicity. Metal-rich environments may also favour the currently preferred model of magnetar progenitors for FRBs via stellar merger events. 

The work by \citet{Grasha2022} and \citet{Ji2022} specifically focused on extragalactic H\,{\sc ii} regions (i.e. ionised atomic hydrogen clouds more common in disc galaxies) in investigating correlations between metallicity and the ionisation parameter. \citet{Mingozzi2020} also highlight that data in their work is dominated by flux from H\,{\sc ii} regions. \citet{Ocker2024} show that the majority of pulsars in the Milky Way with scattering timescales $\tau > 10$\,ms and DM $>$ 600~pc\,cc$^{-1}$ lie behind H\,{\sc ii} regions. \citet{Sicheneder2017} also demonstrated that a single H\,{\sc ii} region along the line of sight to a transient magnetar near Sgr A* can explain the observed pulse broadening.  \citet{Nimmo2025} speculate that a small screen distance measured around FRB\,20221022A is due to the progenitor being embedded in a H\,{\sc ii} region. Given that the spectra in \citet{Gordon2023} used to derive properties such as the mass-weighted age had the slit aligned to cover both the centre of the host galaxy and the FRB localisation,  it could well be that H\,{\sc ii} regions --- either merely in the galaxy disc and along the sightline of the FRB, and/or housing the progenitor itself (potentially a magnetar) --- are a significant contributor to any scattering of FRB pulses.

We refer back to the highly significant result found between scattering timescale and compactness (Section~\ref{sec:compactness}). Tight relations have been found between compactness and gas metallicity \citep{Moran2012,Sanchez2013,Barrera-Ballesteros2016}, where it is proposed that recent stellar mass growth at the edges of galaxies they examined can be linked to the accretion or radial transport of relatively pristine gas from beyond the stellar disc of the galaxy. \citet{Barone2018} and \citet{Barone2020} also found a strong relation between compactness and stellar mass-weighted age. They proposed that galaxies with a higher stellar surface density quench faster and hence earlier, resulting in an older stellar population. This could be why we observe correlations for scattering in compactness, mass-weighted age, and gas-phase metallicity. \citet{Boardman2024} found mass-weighted age to correlate more closely with the gravitational potential $M_{*}$/$R_{\rm eff}$, a result also found by \citet{Sanchez-Menguiano2024}. {\myedit We note that the implication of a connection between increased scattering and an older stellar population is in some tension with theories that H\,{\sc ii} regions (which are associated with younger stars and star formation regions) are a driver of the temporal broadening of FRB pulses \citep{Sicheneder2017}.}

\begin{figure*}
\centering
\resizebox{.95\textwidth}{!}{%
\includegraphics[height=10cm]{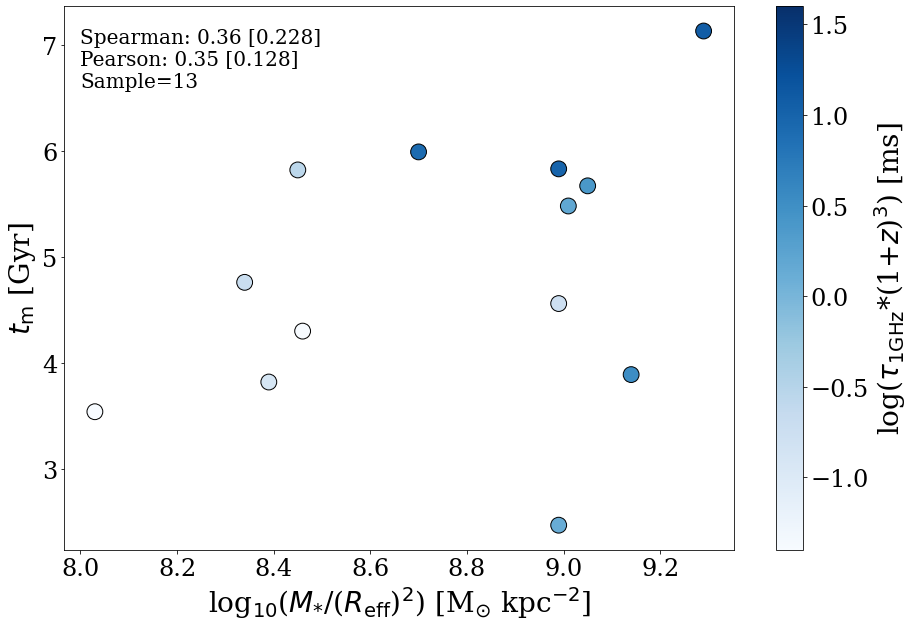}%
\includegraphics[height=10cm]{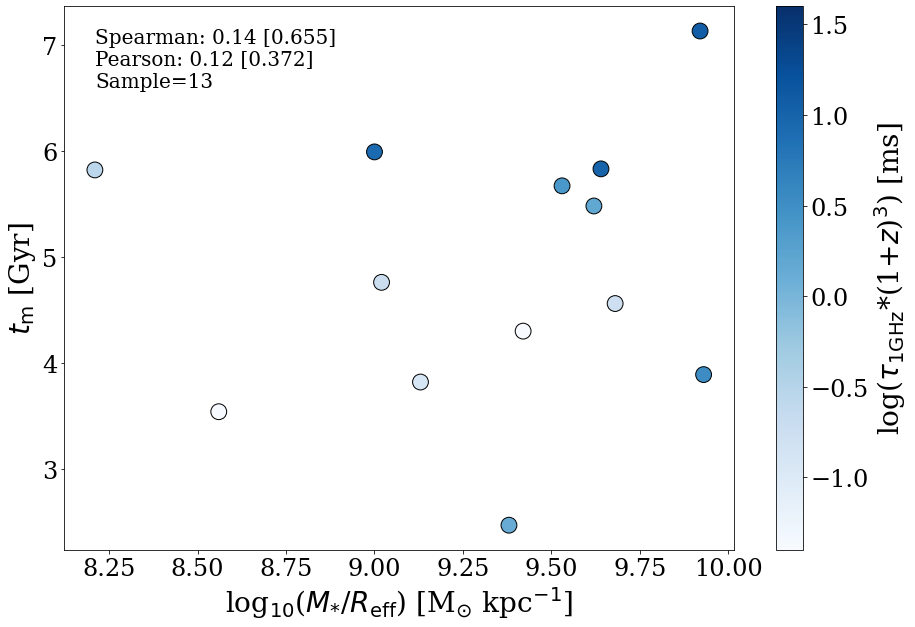}
}
\resizebox{.95\textwidth}{!}{%
\includegraphics[height=10cm]{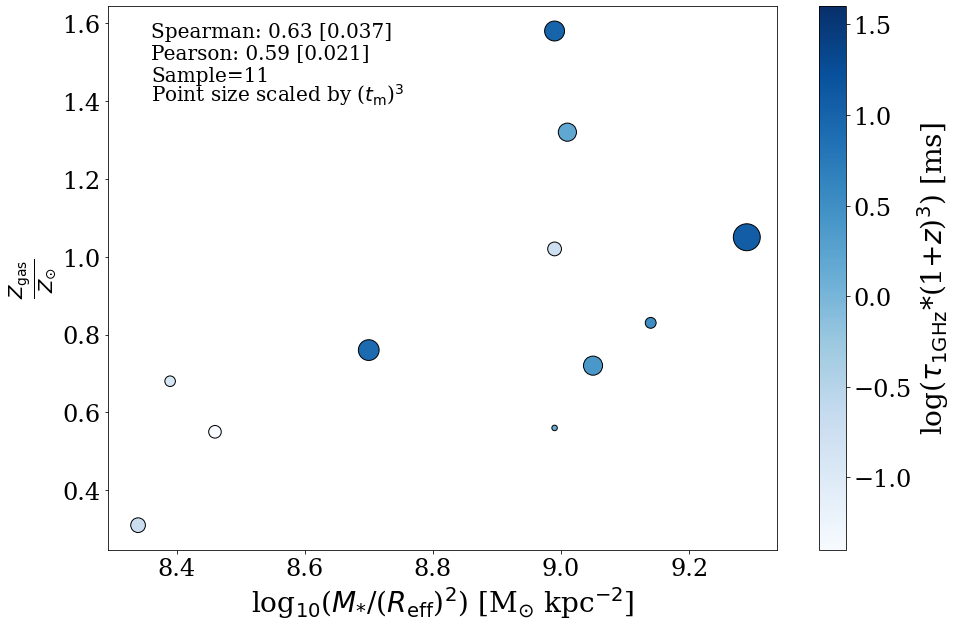}%
\includegraphics[height=10cm]{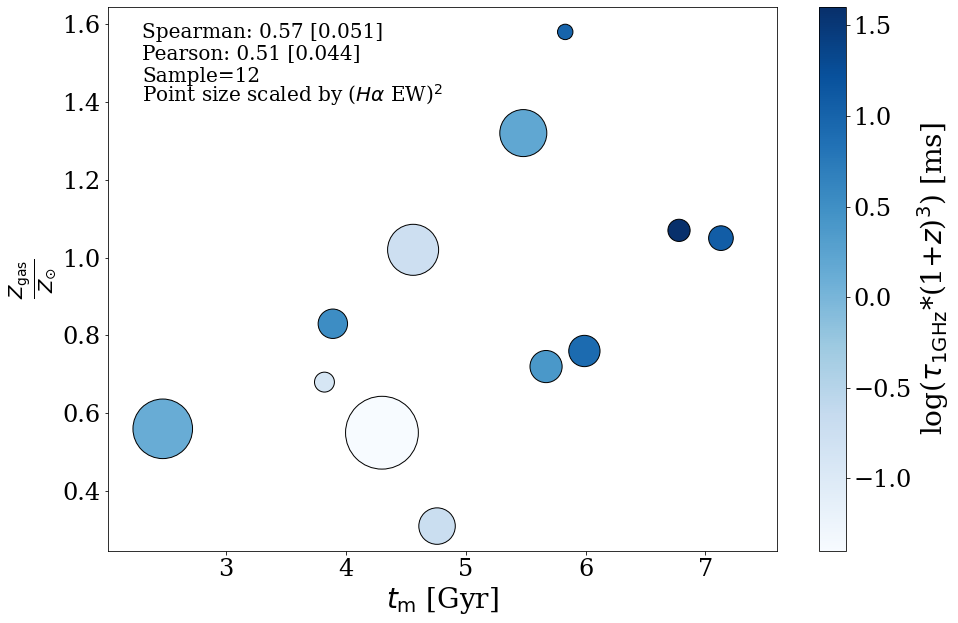}
}
\caption{Comparison of various global galaxy properties of FRB hosts with reliable scattering timescale measures. All points are coloured by the logarithm of the rest-frame scattering timescale at 1~GHz. Spearman and Pearson correlation results are given in the top left as in Fig{\myedit ure}~\ref{fig:taucorr}. Top-left: mass-weighted age $t_{\rm m}$ versus compactness (or stellar surface density $M{*}$/($R_{\rm eff}$)$^{2}$). Top-right: mass-weighted age versus potential $M{*}$/$R_{\rm eff}$. No correlation is seen in either case for the properties compared in the top row. Bottom-left: gas-phase metallicity $\frac{Z_{\rm gas}}{Z_{\odot}}$ versus compactness. In this panel the size of the datapoints scales by a factor of 2$\times$($t_{\rm m}$)$^{3}$. Bottom-right: gas-phase metallicity with host galaxy mass-weighted stellar age. In this panel the size of the datapoints scales by a factor of 2$\times$($H\alpha$EW)$^{2}$. We see correlations for both these relations explored in the bottom row.}
\label{fig:tm_zgas}
\end{figure*}

To investigate if such links hold in our (relatively small) sample of FRB hosts with reliable scattering timescale measurements, we perform the same correlation tests between mass-weighted age with both compactness and potential, and gas-phase metallicity with compactness (top row and bottom-left panel of Fig{\myedit ure}~\ref{fig:tm_zgas}). While we do not see any significant ($p < 0.05$) correlations in the first two cases, we do see correlations in Spearman (0.63; $p = 0.037$) and Pearson (0.59; $p = 0.021$) between gas-phase metallicity and compactness. It is likely that compactness is a key driving factor for the result with $z_{\rm gas}$; more compact galaxies in our sample are found to have a higher gas-phase metallicity, as well as longer scattering timescales. It is however not clear if compactness is the causative observable here, or if alternatively another underlying property is driving these findings. With a larger sample, we might find that a combination of multiple observables such as compactness and gas-phase metallicity provide a stronger correlation than any one property alone.

A weakly significant Pearson anti-correlation (--0.56, $p = 0.025$) is found for $H\alpha$ equivalent width (EW), a measure of current (relative to past) star formation. Bootstrapping suggests that a weak negative correlation coefficient is seen for the Spearman and Pearson statistic 50\% and 53\% of the time respectively, while resampling for errors reproduces a $p < 0.05$ Pearson anti-correlation 49.7\% of the time. This weak negative correlation, if true, does loosely align with the other correlations that a galaxy with older stars and more metal-rich gas leads to greater scattering time, but these parameters are not simple to disentangle (e.g. a galaxy with low $H\alpha$ EW can still have a young mass-weighted age). In Figure~\ref{fig:tm_zgas} (bottom-right panel) we plot mass-weighted age against gas-phase metallicity for our sample where measurements are also available for the scattering time (visualised by the colourbar) and $H\alpha$ EW (visualised by the size of datapoints). While visually lower $H\alpha$ EW (smaller-sized points) is somewhat more typical for galaxies with higher mass-weighted ages, no obvious strong trend is apparent. A Spearman correlation coefficient of 0.57 ($p = 0.051$), and Pearson coefficient of 0.51 ($p = 0.044$) is found between these mass-weighted age and gas-phase metallicity datapoints. This borderline significant correlation is further diminished when considering all available measurements irrespective of whether a robust scattering timescale measurement could be made. A weakly significant Pearson correlation is found between mass-weighted age and $H\alpha$ EW (--0.52; $p = 0.042$) - however this is for a sample of only 12 common datapoints.

When examining the subset of only star-forming hosts (i.e. removing 4 transitioning/quiescent FRB hosts from the sample), for the scattering timescale we see similar or only slightly weakened results for $t_{\rm m}$, $Z_{\rm gas}$/Z$_{\odot}$, and a borderline weak negative Pearson correlation coefficient (-0.66; $p = 0.047$) for $H\alpha$~EW --- the last result may be partly attributed to a smaller sample size (12). In addition, we find a weak negative Pearson correlation coefficient with the $R$-band host galaxy AB magnitude (--0.61; $p = 0.011$), and a Pearson correlation also emerges with $H\alpha$ flux (0.50; $p = 0.011$). By only considering star-forming FRB hosts, it is possible that these trends do slightly impact the measured scattering timescale, but further data is required. %The Pearson correlation strengthens, yet still remains statistically insignificant ($p = 0.150$ for log($M_{*}$)). 

{\myedit We also consider a further subset of FRBs where $\alpha$ values are not consistent with $-4.5 < \alpha < -3.5$. In this case, significant correlations for scattering remain with $Z_{\rm gas}$/Z$_{\odot}$, $H\alpha$\,EW, and compactness, but not mass-weighted age nor potential. We note that in each case only 8--10 datapoints are available which diminishes the ability to establish statistically significant correlations. Other choices of cut in alpha values for the scattering correlation investigation returns similar results.}

\subsection{Rotation Measure correlations}\label{sec:RM}

\begin{figure}
\centering
\includegraphics[width=1.0\textwidth]{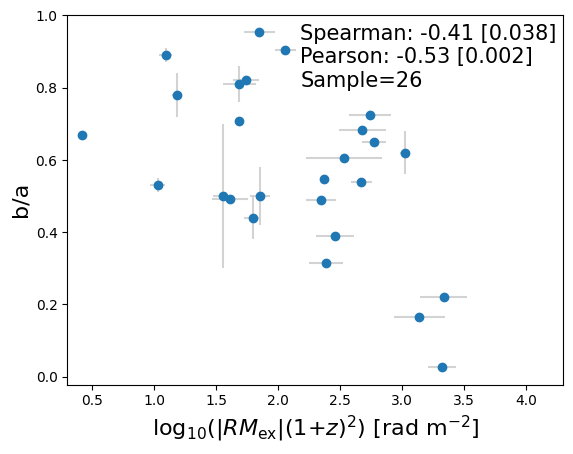}%
\caption{Scatter plot and correlation tests for the rest-frame absolute $RM_{\rm ex}$ of 26 FRBs with the optical disc axis ratio b/a. Smaller values of b/a indicate a more edge-on disc relative to the plane of the sky. Such galaxies may have had the FRB pass through more of the galaxy en route to us and hence increase the rotation measure observed.}
\label{fig:rm_baratio}
\end{figure}

We find one strong anti-correlation with the absolute value of the rest-frame rotation measure, namely with the optical disc axis ratio b/a (Fig{\myedit ure}~\ref{fig:rm_baratio}). We see a Spearman correlation coefficient of --0.41 ($p = 0.038$) and a more statistically significant Pearson correlation of --0.53 ($p = 0.002$); 60.9\% and 82.2\% of our 10,000 bootstrapped results return a Pearson correlation $p < 0.01$ and 0.05 respectively for these two parameters. When incorporating errors and resampling, a $p$-value of $< 0.05$ is recovered for the Pearson correlation 100\% of the time. Smaller b/a values correspond to a more edge-on galaxy, and hence an FRB potentially passing through more of the host ISM is experiencing a longer path length due to the host's magnetic field. {\myedit This aligns with the recent finding by \citet{Khrykin2025}, who measure an average magnetic field strength in the ISM of the FRB host galaxies, and conclude that FRB hosts ``can contribute a non-negligible amount of RM and must be taken into account".} This is an effect only seen for |$RM_{\rm ex}$|; we did not find any correlation for scattering (nor polarisation fraction; Section~\ref{sec:pol}). If the extra RM is coming from an increased path length through a denser medium, then additional DM may also be observed, and a corresponding anti-correlation to exist between b/a and $DM_{\rm host}$. This investigation is left to another work (Marnoch et al., in prep.). With a larger sample one could also place constraints on the contributions of the host galaxy  and progenitor environment to RM when accounting for the b/a of the host galaxy. This result is coupled with a weak positive Pearson correlation (0.37; $p = 0.040$) with galaxy inclination angle, which is inferred from the disc axis ratio b/a but relies on assumptions about the relative bulge size and disc thickness. However, bootstrapping suggests that only $40.2$\% of the time a weak ($p < 0.05$) Pearson correlation will be found with inclination angle. % or the cosine of the inclination respectively. 

We do not find any other correlations with any global galaxy property for |$RM_{\rm ex}$| - a weakly significant Pearson correlation ($p$ = 0.033) was seen for the \textit{linear} measure of compactness in a sample of 16, but not the logarithmic values presented in Table~\ref{tab:correlations}. The lack of significant correlations (or anti-correlations) with other host galaxy properties may inform models on the FRB progenitor, e.g. that the local environment of the progenitor is significant. %, and that any other property of the FRB in the host galaxy ISM does not significantly impact the observed RM of the FRB. 
However, care must be taken in regards to the optical disc of the host with |$RM_{\rm ex}$| measurements attributed to the progenitor environment, particularly for FRBs without a large offset from the optical galaxy centre. We again stress that this is an initial study that would benefit from larger sample sizes. These are also rotation measure values for apparently one-off FRBs, while it is documented that repeating FRBs can have significant changes in the recorded $RM$, even to the point of magnetic field reversals \citep{Anna-Thomas2023} - another indicator on the importance of the progenitor environment on observed $RM$ of FRB profiles. 

\subsection{Polarisation fraction correlations}\label{sec:pol}

%In some cases, the number of FRBs with a satisfactory number of circular or linear polarisation fractions for a given host galaxy property is lacking (e.g. sample of merely 9 for $H\alpha$ flux and EW). For such properties, we recognise that further statistics are necessary for higher confidence results. 
%NOW: our lowerest sample size is 14, don't feel this is necessary now. but can re-include and change 9 -> 14 or 15 in some cases 

\begin{figure}
\centering
\includegraphics[width=1.0\textwidth]{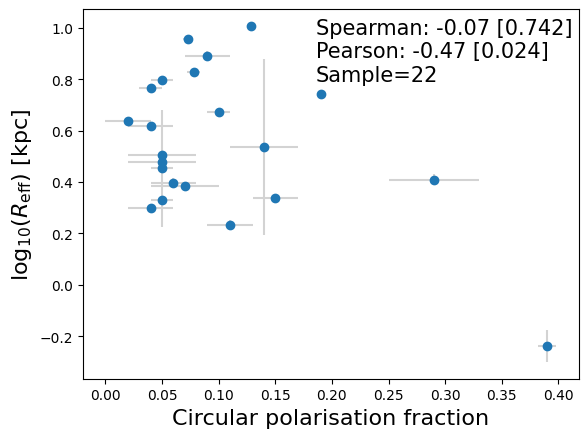}%
\caption{Scatter plot and correlation tests for circular polarisation fraction of 27 FRBs with the logarithm of the effective radius of the host galaxy. If the lower-right datapoint is excluded (FRB\,20230708), no significant negative correlation is found.}
\label{fig:circ_frac_reff}
\end{figure}

Amongst the correlation tests for circular polarisation fraction is the result for the (logarithm of the) %offset of the FRB localisation from the centre of the host galaxy
effective radius of the galaxy (Fig~\ref{fig:circ_frac_reff}); a negative Pearson correlation coefficient of -0.47 ($p = 0.024$) is found for a sample of 22 FRBs and their hosts. %(For the non-log meausre of $R_{\rm eff}$, a p-value of 0.052 is found for Pearson correlation coefficient of --0.38). 
%NO SCATTER UPPER LIMS: negative Spearman and Pearson correlations of --0.58 ($p = 0.014$) and -0.62 ($p = 0.002$) respectively are found for a sample of 18 FRB host galaxies. 
%The further away from the centre of the galaxy an FRB may arise in, the lower the circular polarisation. Therefore, it is possible that any measured circular polarisation for an FRB may be attributed to the host galaxy if it is more centralised. 
The larger the host galaxy, the more content the FRB may be passing through which could impact circular polarisation upon the pulse. From bootstrapping this weak negative Pearson correlation is seen 56.0\% of the time. 
This is tempered by the fact we do not know if an FRB occurs on the near or far side (or in the middle) of the galactic disc with respect to us. This correlation is also not significant when considering the linear measure of the effective radius. 

This correlation is driven by FRB\,20230708A (lower-right datapoint in Figure~\ref{fig:circ_frac_reff}), an FRB with rich intrinsic temporal structure suggestive of quasi-periodicity \citep{Dial2025}. This source has the highest circular polarisation fraction, and the smallest host galaxy of our sample with polarisation fraction measures \citep{Shannon2024}. This host is the least-luminous yet found for a non-repeating FRB \citep{Muller2025}. \citet{Dial2025} did not find any evidence that the circular polarisation of FRB\,20230708A was due to Faraday conversion. If we exclude FRB~20230708A, no significant correlation is found for circular polarisation fraction with the effective radius, nor its logarithm ($p > 0.4$). \citet{Scott2025} highlights a break in the cumulative distribution of circular polarisation fraction for both CRAFT and DSA-110 FRBs at Stokes $V > 20$\%, which suggests a distinct sub-class of highly circularly polarised FRBs, consisting of $\sim$10\% of the FRB population.
Therefore we do not believe this correlation result is significant across the whole population of FRB hosts considered in this work. 
A more nuanced study into the position of an FRB with respect to star formation in spiral arms of the host and their polarisation fractions (Mannings et al., in prep.) may shed additional light on this tenuous correlation \citep[see also the study by][highlighting observed circular polarisation in star-forming regions]{Bailey1998}.
A similar result is seen with circular polarisation and stellar mass, which is again dependent on the inclusion of FRB\,20230708A (that is, no correlation is seen when excluding this FRB and its host). While we see weakly significant Pearson anti-correlations for the circular polarisation fraction with linear measures of compactness and potential, these results require omitting FRB\,20230708A and are not significant for the logarithmic galaxy measures. % Conversely, 
%a weak correlation with a negative trend between the circular polarisation fraction and the stellar mass-weighted age (Spearman correlation coefficient of --0.61; $p = 0.026$) is seen when \textit{not} including FRB\,20230708A; no significant Spearman correlation ($p = 0.235$) is seen when including the mass-weighted age for FRB\,20230708A (Muller et al., in prep). %No obvious outlier point is evident in Figure~\ref{fig:circ_frac_tm} for this comparison, and FRB\,20230708A is not included in this result (as it has no mass-weighted age measurement). 
%From bootstrapping, this weak negative correlation without FRB\,20230708A for is seen 62.1\% of the time. %Similarly, 
%a Pearson anti-correlation is seen for the linear measure of potential (--0.43; $p = 0.032$), and borderline insignificant anti-correlation with linear measure of compactness (Spearman: --0.52, $p = 0.069$; Pearson: --0.4, $p = 0.052$). We note these are not seen for the logarithmic measures of potential and compactness. %A borderline insignificant Spearman correlation is also seen with the mass-weighted age (--0.51; $p = 0.060$). Host galaxies with younger stellar populations or a higher potential/are more compact and may hence impart a higher circular polarisation fraction on the FRB signal. However, this requires removing one datapoint (FRB\,20230708A); see also Section~\ref{sec:caveats}.

%newer result involving 230708A rules this out! Might be an outlier point but was a weaker trend
%\begin{figure}
%\centering
%\includegraphics[width=1.0\textwidth]{figures/Circular_polarisation_fraction_tm_[Gyr].png}%
%\caption{Scatter plot and correlation tests for circular polarisation fraction of 15 FRBs with the stellar mass-weighted age of the FRB host galaxy.}
%\label{fig:circ_frac_tm}
%\end{figure}

No strongly significant correlations are found for the listed parameters in Table~\ref{tab:correlations} with linear polarisation fraction. %Two weak positive Pearson correlations are seen for the linear polarisation fraction: with the host galaxy's AB magnitude, and the $H\alpha$ flux. However, bootstrapping suggests this weak correlation is significant less than half the time. 
As with circular polarisation fractions, some weakly significant results are seen for linear measures of SFR and sSFR, which we attribute to one datapoint: that of FRB\,20211127I, which has a remarkably high SFR of $>35 M_{\odot}$ \citep{Glowacki2023,Gordon2023}, much greater than any other FRB host considered here. Excluding this FRB from the sample removes any significant result. It does not drive any other correlation result found such as for the scattering timescale discussed in Section~\ref{sec:scatter}. 
%We highlight that for this sample we also see a weak correlation with the \textit{linear} SFR and sSFR (rather than the logarithmic measures we present in Table~\ref{tab:correlations}). This is also driven by one datapoint; that of FRB\,20211127I, which has a remarkably high SFR of $>35 M_{\odot}$ \citep{Glowacki2023,Gordon2023}, much greater than any other FRB host considered here. If we excluded FRB\,20211127I from the sample, such correlations with linear polarisation fraction are no longer significant ($p > 0.2)$. This FRB does not drive any other correlation result found such as for the scattering timescale discussed in Section~\ref{sec:scatter}. We hence do not find these correlations with linear polarisation fraction to be currently significant, but a larger sample size with a more diverse range of FRB hosts should be investigated. 
The same findings are seen for the total polarisation fraction, which is typically dominated by the linear polarisation fraction in this sample. 
%A strong negative Spearman correlation between the linear polarisation fraction and dust extinction by old stars (--0.79; $p = 0.005$) is observed, albeit for a sample of only 9. This also sees a weak negative Pearson correlation (--0.73; $p = 0.026$). This dust extinction effect may hence be important to consider in linear polarisation fraction studies of FRBs. %A stronger positive correlation (Spearman of 0.83; $p = 0.010$, and Pearson of 0.75; $p = 0.032$) with stellar metallicity is also seen for a sample of 8 - ergo, any measures of the linear polarisation fraction for FRBs with a high stellar metallicity should consider this preliminary finding. 
%This is similarly seen for the total polarisation fraction, as expected as most of the CRAFT ICS FRBs are dominated by linearly polarised FRBs (all but one FRB considered here have circular polarisation fractions $<$~0.2).
%NO LONGER - unless we exclude 211127, and not others with poor scattering fits. ergo, don't think this is significant especially with now larger sample size. 
No different correlation result arises for polarisation fraction nor RM when considering only star-forming FRB host galaxies. 

\subsection{Significance of results}\label{sec:caveats}

In all, 21 galaxy properties were used for this study (20 when considering only the linear or log measure of effective radius).  Taking the number of 20 galaxy properties we could naively expect a weakly significant ($p < 0.05$) correlation to arise purely from random chance per Spearman and Pearson correlation, for each of the scattering timescale, |$RM_{\rm ex}$|, and circular/linear polarisation FRB properties, and 1 strongly significant ($p < 0.01$) event overall. This assumption depends on all galaxy quantities are independent of each other; see below. 

To better quantify this, we randomly resample each dataset so a rest-frame $\tau$ measurement is paired with a random host galaxy property, compute the resulting Spearman and Pearson correlation for the shuffled dataset, and repeat 10,000 times. We then find how often a weakly or strongly significant correlation arises from these shuffled datasets (Table~\ref{tab:shufflefraction}). More than 3 Spearman correlations with $p \leq 0.05$ occur 2.4\% of the time, and more than 2 correlations where $p \leq 0.01$ occur 0.2\% of the time (exactly 2 strongly significant correlations 2\% of the time). Likewise for Pearson, more than 4 correlations with $p \leq 0.05$ occur 3.9\% of the time, and more than 1 correlation with $p \leq 0.01$ 6.3\% of the time (exactly 1 strongly significant correlation occurs 29.1\% of the time). We remind the reader that without reshuffling scattering-based datasets, we found 3 [2] Spearman correlations with $p < 0.05$ [0.01], and 5 [1] Pearson correlations with $p < 0.05$ [0.01] respectively (Table~\ref{tab:correlations}). 

We find similar results for the rest-frame |$RM_{\rm ex}$| when shuffling randomly, and again when combining both scattering and |RM| measurements. For the `true' datasets for |RM| we found 1 Spearman correlation with $p < 0.05$ and 2 [1] Pearson results with $p < 0.05$ [0.01]. Correspondingly, by randomly shuffling our Spearman result would be expected to happen 38.1\% of the time, and one or less strongly significant Pearson correlation to occur 91\% of the time (more than one 9\% of the time). We conclude that we do not expect high numbers of significant correlations, but that it is also unlikely that all correlations presented here are real based on these assumptions.

\begin{table}[h!]
\centering
\caption{Fraction of randomly shuffled correlations with $N$ strongly or weakly significant results ($p \leq 0.01$ or 0.05 respectively), where an FRB pulse property has been randomly paired with one of the global galaxy properties considered in this study. We list fractions for correlations between just scattering timescale, just absolute RM, and both.}
\addtolength{\tabcolsep}{-0.2em}
\begin{tabular}{llllll}
\toprule
 %FRB measure & Galaxy measure & Fraction Spearman $p < 0.01$ & Fraction Spearman $p < 0.05$ & Fraction Pearson $p < 0.01$ & Fraction Pearson $p < 0.05$ \\
  FRB  & $N$ & Fraction & Fraction  & Fraction & Fraction  \\
 measure & & Spearman & Spearman &  Pearson & Pearson \\
 &  & $p \leq 0.01$ & $p \leq 0.05$ &  $p \leq 0.01$ & $p \leq 0.05$ \\
\midrule
 log($\tau_{\rm 1 GHz}$*(1+$z$)$^{3}$) & 0 & 0.776  & 0.311  & 0.646 & 0.106 \\
 & 1 & 0.202 & 0.388  & 0.291 & 0.233  \\
 & 2 &  0.020  & 0.211 & 0.054 & 0.303  \\
 & 3 & 0.002  & 0.066  & 0.007  & 0.211  \\
 & $\geq$4 & 0.000 & 0.024 & 0.002 & 0.147 \\
 \midrule
  log(|$RM_{\rm ex}$|*(1+$z$)$^{2}$) & 0 & 0.782  & 0.335  & 0.593  & 0.103  \\
 & 1 & 0.201  & 0.381  & 0.317  & 0.251 \\
 & 2 & 0.015  & 0.195  & 0.076  & 0.279 \\
 & 3 & 0.002  & 0.073 & 0.014  & 0.201 \\
 & $\geq$4 & 0.000 & 0.016 & 0.000 & 0.166 \\
  \midrule
  Both $\tau$ and |$RM$| & 0 & 0.779  & 0.323  & 0.620  & 0.105  \\
 & 1 & 0.202  & 0.3851  & 0.304  & 0.242 \\
 & 2 & 0.018  & 0.203  & 0.065  & 0.291 \\
 & 3 & 0.002  & 0.070 & 0.010  & 0.206 \\
 & $\geq$4 & 0.000 & 0.020 & 0.001 & 0.157 \\
  \bottomrule
\end{tabular}
\label{tab:shufflefraction}
\end{table}

We note that not all of these galaxy measurements are independent of each other - some measures are combinations of two other properties (such as sSFR or compactness), while others are derived from the other (e.g., inclination is based on the optical disc axis ratio b/a). As discussed earlier, correlations have been found between various measures (e.g. compactness and potential with mass-weighted age and metallicity) in the literature, which we find also correlate with scattering {\myedit in our FRB sample (Figure~\ref{fig:taucorr}).} Nor do we have reason to believe that some properties such as dust extinction in the host galaxy or its $R$-band magnitude would have significantly driven a relation in (e.g.) scattering timescale prior to carrying out this analysis. Scattering of FRBs should be driven by both the density of ionised electrons and turbulence, and some global galaxy properties are less likely to be dependent on those aspects. As only some of these galaxy properties or measurements could be considered independent of others, others that have been shown to correlate with each other and with $\tau$ may be a result of an underlying property manifesting as a partial correlation we find in each case. %{\myedit Galaxy properties that are dependent on each other may also reflect in the correlations found with FRB properties. If by random chance a correlation is found with a galaxy property, then other dependent properties are also expected to be found by chance.}
%Galaxy properties that are dependent on each other may also reflect in the correlations found with FRB properties. If by random chance a correlation is found with a galaxy property, then other dependent properties are also expected to be found by chance.
%A more conservative estimate of 10 unique galaxy property measurements lowers the expected number of significant correlations to arise from random chance. 
 %- I agree with the sentiment, but doesn't that assume that you are able to identify - and test against - these 10 unique properties, rather than the larger number of partially correlated ones that are currently being used? Say there are really 10 independent properties, that can be munged into 20 different partially correlated properties. I think the number of times you would see >N chance correlations is smallest if you test against the 10 independent quantities), it would get larger if you tested against 20 independent quantities, but it would be largest of all if you test against 20 partially correlated quantities (since if there *is* a chance correlation with one of the truly independent underlying properties, it is likely to manifest in more than one of the partially correlated observables). So I don't think I agree with this sentence as currently written.

We do see a few weakly significant correlations, and so each of those can have reasonable doubt placed upon them. In particular, the results discussed for circular and linear polarisation fraction (the former dependent on the inclusion or exclusion of FRB\,20230708A) can be potentially dismissed given the current sample of 14 or 15 datapoints. We would however be particularly `unlucky' to see all three strong ($p < 0.01$) correlations to have arisen by chance, particularly the correlation between scattering timescale and compactness ($p = 0.001$), and the anti-correlation found between |$RM_{\rm ex}$| and b/a ($p = 0.002$). We also do not expect a `perfect' correlation to arise with global galaxy properties, as we generally expect the immediate environment of the FRB progenitor, likely not entirely dependent on the host galaxy, to contribute to scattering for at least some FRBs. Factoring this in however requires both very high accuracy localisation and dedicated follow-up studies of the host. 

\begin{figure}
\centering
\includegraphics[width=1.0\textwidth]{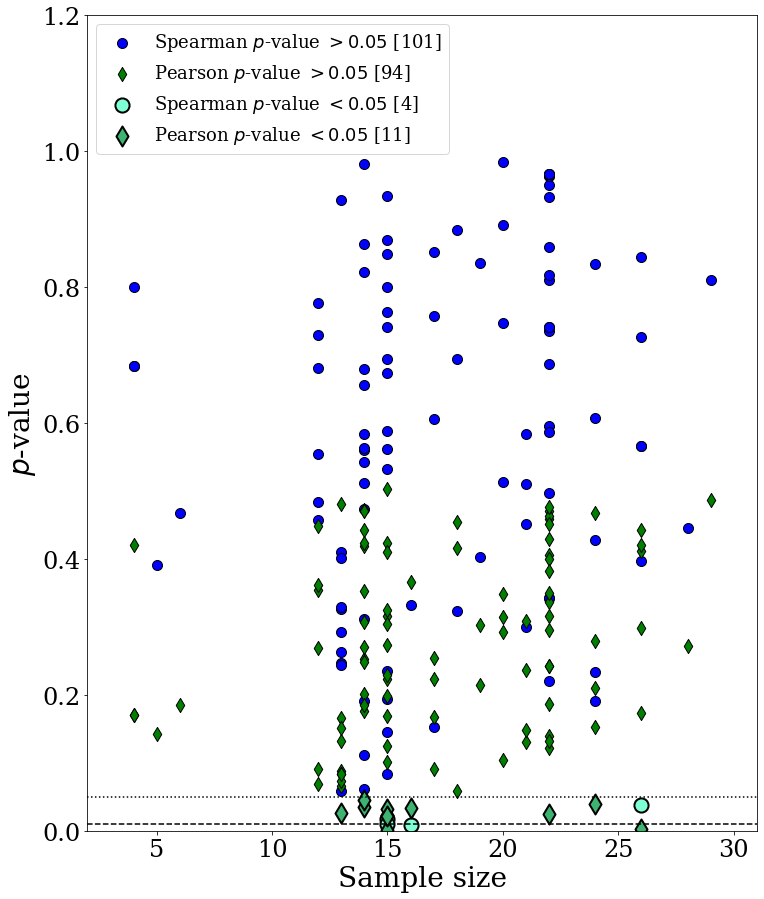}%
\caption{Comparison of $p$-values found for Spearman and Pearson correlation tests versus sample size. Dashed horizontal lines indicate $p$ of 0.01 and 0.05.}
\label{fig:pval_n}
\end{figure}

In Fig.~\ref{fig:pval_n} we show the distribution of $p$-values found versus sample size as given in Table~\ref{tab:correlations}. We find 15 cases where $p$ indicates a statistically significant correlation, and see a spread in sample size. This is slightly above expectations for the number of $p$ to fall below 0.05 ($\sim10$), assuming completely independent properties had been tested; three cases have $p < 0.01$ (2 with $p \leq 0.003$), unlikely to all happen by random chance. This includes the results for circular polarisation fraction which were sometimes dominated by one FRB, but not the linear polarisation fraction trends with linear SFR and sSFR, which were also dominated by one FRB with an outlier SFR (Section~\ref{sec:pol}). 

We stress that the sample size of this study is still fairly modest, and so these are merely plausible explanations for the correlations observed here, assuming they are real. Larger samples of well-localised FRBs with reliable scattering-time measurements \textit{and} host galaxy property studies are necessary to further investigate these findings, and extend the study across a range of galaxy morphologies and star formation rates, and FRB populations (e.g. apparent one-off vs repeating bursts). A better understanding of the gas fraction of FRB host galaxies, such as through studies of the neutral hydrogen (H\,{\sc i}) content \citep[e.g.][Roxburgh et al. in prep.]{Kaur2022, Glowacki2023, LeeWaddell2023} would be potentially useful in determining if these correlations are stronger in host galaxies with higher gas fractions, i.e. larger reservoirs of star-forming gas that could lead to the birth of FRB progenitors, or H\,{\sc ii} regions.

\section{Conclusion}\label{sec:conclusion}

We present the first investigation of correlations between multiple FRB properties (scattering timescales, polarisation fractions, and extragalactic rotation measure) with a range of global host galaxy properties. 

Our investigation into the scattering timescales produces a few notable statistically significant positive correlations: with the stellar surface density or compactness, the stellar mass-weighted age, and the gas-phase metallicity. A less significant correlation with the gravitational potential, and an anti-correlation with $H\alpha$ EW measures are also found. An FRB travelling through a more compact (i.e. dense) host galaxy may hence undergo more scattering, and we also find a correlation between compactness and gas-phase metallicity for our FRB hosts. Alternatively, a higher gas metallicity in the ISM of the host galaxy can lead to a harder radiation field, and increase in the ionised baryon fraction, which would contribute to an increased scattering time of the FRB. It appears that FRB scattering is correlated with some property (or multiple) of the host galaxies, but it is not clear which of our observables is the causative one, or if none of them are and they are all related to some other underlying property of compactness, mass-weighted age, and gas-phase metallicity.

%We also postulate that a higher gas metallicity in the ISM of the host galaxy can lead to a harder radiation field, and increase in the ionised baryon fraction, which would contribute to an increased scattering time of the FRB, rather than the local environment of the progenitor being a driving factor of all scattering. H\,{\sc ii} regions may be a contributor, which either lie along the line-of-sight of the FRB within the typically star-forming hosts of FRBs, and/or potentially host the FRB progenitor itself. 
We highlight the lack of any scattering timescale correlation with the galaxy inclination angle or optical disc axis ratio, in tension with the study by \citet{Bhardwaj2024}, which found an inclination-related bias against detecting FRBs in the most edge-on host galaxies. For the majority of other host galaxy properties, we find no correlation with scattering. 

We also find a strong Pearson correlation between the absolute value of the extragalactic RM with the optical disc axis ratio b/a - i.e. more edge-on discs correspond to a higher FRB RM. The absence of correlations of other global galaxy properties with rotation measure, besides galaxy inclination which is derived from b/a, suggests that the local FRB progenitor environment may be more significant than the host galaxy average values, but it appears the orientation of the host galaxy disc should be considered. We find a few weak correlations for circular polarisation fraction but conclude they are driven by FRB\,20230708A, and see no significant correlations when excluding this datapoint, rather than such results holding for the rest of the sample analysed here. As the sample sizes are still relatively modest, we highlight some caution in interpreting all these results, but find it unlikely that all have arisen by random chance. 
%couple interesting negative correlations are found for circular polarisation fraction (with the effective radius of the gost galaxy) and linear polarisation (with dust extinction by old stars), but these come with the caveat of low ($<15$ or even $<10$) sample sizes. 

We encourage further follow-up studies of well-localised FRBs to expand this initial finding to a larger sample size. Given the advances by FRB survey teams and new radio telescopes, and the potential for many more host galaxy property measurements in coming years, this may not be too far off. %Additionally, ensuring that we are not merely sensitive to the effect of e.g. H{\sc ii} regions that are housing the FRB progenitor rather than within the host galaxy and along the FRB sightline is vital.

\begin{acknowledgement}
{\myedit We thank the anonymous referee for useful feedback that has improved this manuscript.} We thank Nicholas Boardman, Romeel Davé, Ron Ekers, Sergey Koposov, Avery Meiksin, Michael Petersen, and Dirk Scholte for useful discussions and suggestions that have contributed to the findings of the paper. 

This work was performed on the OzSTAR national facility at Swinburne University of Technology. The OzSTAR program receives funding in part from the Astronomy National Collaborative Research Infrastructure Strategy (NCRIS) allocation provided by the Australian Government, and from the Victorian Higher Education State Investment Fund (VHESIF) provided by the Victorian Government.

This work was supported by software support resources awarded under the Astronomy Data and Computing Services (ADACS) Merit Allocation Program. ADACS is funded from the Astronomy National Collaborative Research Infrastructure Strategy (NCRIS) allocation provided by the Australian Government and managed by Astronomy Australia Limited (AAL).

This scientific work uses data obtained from Inyarrimanha Ilgari Bundara, the CSIRO Murchison Radio-astronomy Observatory. We acknowledge the Wajarri Yamaji People as the Traditional Owners and native title holders of the Observatory site. CSIRO’s ASKAP radio telescope is part of the Australia Telescope National Facility (https://ror.org/05qajvd42). Operation of ASKAP is funded by the Australian Government with support from the National Collaborative Research Infrastructure Strategy. ASKAP uses the resources of the Pawsey Supercomputing Research Centre. Establishment of ASKAP, Inyarrimanha Ilgari Bundara, the CSIRO Murchison Radio-astronomy Observatory and the Pawsey Supercomputing Research Centre are initiatives of the Australian Government, with support from the Government of Western Australia and the Science and Industry Endowment Fund.

This research has made use of NASA's Astrophysics Data System Bibliographic Services. This work is based on observations collected at the European Southern Observatory under programme IDs 0103.A-0101, 105.204W, and 108.21ZF.

\end{acknowledgement}

\paragraph{Funding Statement}

MG and CWJ acknowledges support by the Australian Government through the Australian Research Council Discovery Projects funding scheme (project DP210102103). MG also acknowledges support through UK STFC Grant ST/Y001117/1. MG acknowledges support from the Inter-University Institute for Data Intensive Astronomy (IDIA). IDIA is a partnership of the University of Cape Town, the University of Pretoria and the University of the Western Cape. For the purpose of open access, the author has applied a Creative Commons Attribution (CC BY) licence to any Author Accepted Manuscript version arising from this submission.
ACG, ARM, JXP, %WF, RAJ, 
and NT\ acknowledge support from NSF grants AST-1911140, AST-1910471 and AST-2206490 as members of the Fast and Fortunate for FRB Follow-up team. RMS acknowledges support through ARC Discovery Project DP220102305. ARM acknowledges support from the National Science Foundation under grant AST-2206492 and from the Nantucket Maria Mitchell Association. ACG and the Fong Group at Northwestern acknowledge support by the National Science Foundation under grant Nos. AST-1909358, AST-2206494, AST-2308182 and CAREER grant No. AST-2047919.

\paragraph{Competing Interests}

None.

%A statement about any financial, professional, contractual or personal relationships or situations that could be perceived to impact the presentation of the work --- or `None' if none exist.

\paragraph{Data Availability Statement}

Data used in this study is presented in Tables~\ref{tab:burstprop},~\ref{tab:galprop}, and~\ref{tab:galfitprop}. These values can be primarily found in \citet{Scott2025} and \citet{Gordon2023}, and elsewhere in the literature as noted in the table captions. \citet{Muller2025} will also contain FRB host data used in this study. Other relevant FRB data can be made available upon reasonable request to the CRAFT team. 
%A statement about how to access data, code and other materials allowing users to understand, verify and replicate findings --- e.g. Replication data and code can be found in Harvard Dataverse: \verb+\url{https://doi.org/link}+.

%\endnote in some journals will behave like \footnote; and \printendnotes will not output anything. 
\printendnotes

%\printbibliography
\bibliography{main.bib}

%\appendix

%\section{Example Appendix Section}

\end{document}